\documentclass[%
prx,
reprint,
superscriptaddress,
nofootinbib,
 amsmath,amssymb,
 aps,
floatfix,
]{revtex4-2}

\usepackage{graphicx}
\graphicspath{
{./figures/}
{./figures-SM}
{./figures-storage}
}

\usepackage[caption=false]{subfig}

\usepackage{lineno}
\modulolinenumbers[5]
\let\oldalign\align
\let\oldendalign\endalign
\renewenvironment{align}{%
    \linenomathNonumbers\oldalign%
    }{%
    \oldendalign\endlinenomath%
    }

\let\oldequation\equation
\let\oldendequation\endequation
\renewenvironment{equation}{%
    \linenomathNonumbers\oldequation
    }{%
    \oldendequation\endlinenomath%
    }

\usepackage{graphicx}% Include figure files
\usepackage{dcolumn}% Align table columns on decimal point
\usepackage{bm}% bold math
\usepackage{bbold}%blackboard bold font
\usepackage{bbm}% blackboard bold font (matching a font available in Matplotlib)
\usepackage[dvipsnames]{xcolor}
\usepackage{scrextend}%Repeat a footnote multiple times
\usepackage{multirow}
\usepackage{makecell}%For newlines within table cells
\usepackage{relsize}%Larger integrals

\def\change#1{{#1}}

\definecolor{redcolor}{rgb}{.7,0.,0.}

\definecolor{jmmcolor}{rgb}{0.0, 0.7, 0.0}

\renewcommand{\arraystretch}{1.3}%More space between rows of table

\newcommand{\xin}{{{\bf x}^{\text{input}}}}

\newcommand{\xvecit}[1]{{\bf x}^{(#1)}}
\newcommand{\xit}[1]{{x^{(#1)}}}

\newcommand{\pp}[2]{\frac{\partial #1}{\partial #2}}

\newcommand{\bbeta}{\boldsymbol{\beta}}

\newcommand{\vecto}[3]{\left( \begin{array}{c} #1 \\ #2 \\ #3 \end{array} \right)}%A command for column 3-vectors.

\begin{document}

% \title{Distributional-constrained surrogates}
%\title{Constrained-likelihood surrogates for estimation and hypothesis testing}
% \title{Constrained surrogates for arbitrary probability distributions}
\title{Constrained surrogates for arbitrary families of continuous probability distributions
}

\author{Jack Murdoch Moore}
\email{jackmoore@tongji.edu.cn}
\affiliation{%
MOE Key Laboratory of Advanced Micro-Structured Materials, and School of Physical Science and Engineering, Tongji University, Shanghai 200092, People’s Republic of China
}%
\affiliation{%
National Key Laboratory of Autonomous Intelligent Unmanned Systems,
MOE Frontiers Science Center for Intelligent Autonomous Systems, Tongji University,
Shanghai 200092, People’s Republic of China
 }%

\author{Eduardo G. Altmann}%
\email{Contact author: eduardo.altmann@sydney.edu.au}%
 \affiliation{%
 School of Mathematics and Statistics, University of Sydney, New South Wales 2006, Australia
 }%

\date{\today}% It is always \today, today,

\begin{abstract}
Empirical regularities described by parametrized families of probability distributions can shape how complex systems are interpreted and explained. A recurring problem is to determine the extent to which these descriptions are supported by observations, and how they can be used to characterize the underlying system.
Here, we introduce a general framework for constructing constrained surrogates capable of representing arbitrary families of continuous distributions. We demonstrate its utility by providing explicit algorithms for several canonical distributions in statistical physics, including continuous powerlaw, lognormal, exponential, and truncated Gaussian distributions. These constrained surrogates enable unbiased estimation and accurate tests of model validity based on arbitrary (physically meaningful) sample statistics, without requiring specification or estimation of model parameters. We show how these theoretical advantages translate into more informative and nuanced conclusions in applications ranging from bushfire magnitudes and word usage in texts to brain connectomes and city populations.
\end{abstract}

\maketitle

%\tableofcontents

\section{Introduction}

Computation enables exploration of complex systems that are inaccessible to purely analytic approaches or direct experimentation. For this reason, it has come to stand alongside theory and experiment as a third pillar of science~\cite{rude2018research,kovalchuk2020twenty}. A computational strategy with close ties to both theory and experiment is the method of surrogates, which generates synthetic datasets---known as surrogates---by imposing a theoretical hypothesis or model on empirical data~\cite{theiler1992testing,theiler1996constrained,lancaster2018surrogate}. Surrogates can be used to estimate properties of observed systems~\cite{clauset2013estimating,pasquini2018first}, test model validity~\cite{clauset2009power,deluca2013fitting}, and compare candidate models~\cite{malevergne2011testing,papana2021detecting}. As a result, surrogate methods are applied across scientific domains, including neuroscience~\cite{antonacci2024spectral}, climate science~\cite{donges2009backbone}, epidemiology~\cite{scarpino2019predictability}, and artificial intelligence~\cite{zhang2025brain}.

Both the theoretical support and the performance of surrogate methods depend on how they are constructed. Surrogates methods relying on a fully specified generative model risk bias when the assumed model is misspecified~\cite{theiler1996constrained,lancaster2018surrogate,moore2022nonparametric}. Constrained surrogates address this limitation by avoiding explicit specification of a single generative process and instead constraining quantities observed in the data whose preservation allows representation of the relevant class of models in generality%
~\cite{%
kugiumtzis1999test,%Comparison of AAFT and IAAFT surrogates
% kreuz2004measure,%PRE: Constrained surrogates which preserve seizure times and some other properties, for validating seizure prediction methods
% keylock2006constrained,%PRE: Constrained time-series surrogates based on wavelets and designed to preserve time and frequency properties
% ansmann2011constrained,%PRE: Constrained randomization of weighted networks preserving weight sequence or node strength sequence
lucio2012improvements,%PRE: Subtract trend, apply constrained surrogatea algorithm, add trend back
% besag2013exact,%Constrained surrogates for Markov order
pethel2014exact,%Physica D: Exact hypothesis tests using Markov-order constrained surrogates
% gonzalezcastillo2015tracking,%PNAS: Apply phase-randomized time series constrained surrogates and link-weight randomisation when studying whole-brain fMRI dynamics
% suweis2015effect%Degree-constrained randomised networks
% yeshurun2017amplification,%PNAS: Constrained surrogates from randomly reordering words in sentences
% papana2017assessment,%Use time-shifted surrogates to assess significance of estimated causal strength
% alexander2018testing,%NeuroImage: Propose constrained "spatial permutation framework to generate null models of overlap"
siggiridou2019evaluation,%Use time-shifted surrogates to compare significance of Granger causality measures for constructing networks from multivariate time series
correa2020constrained,%Physica D: Exact hypothesis tests using MCMC-based Markov-order constrained surrogates
stramaglia2021local,%PRE: Uses IAAFT surrogates to assess significance of local Granger causality
%
% deco2021revisiting,%Nat. Hum. Behav.: Uses classical constrained surrogates to investigate brain dynamics
markello2022neuromaps,%Nat. Methods: Apply a constrained "spatial permutation framework to generate null models of overlap"
vasa2022null,%Nat. Rev. Neurosci.: Review of constrained null and surrogate models in network neuroscience
ceolini2024age,%PNAS: Applies shuffling and some type of FT surrogates to behavioral time series
% pinto2024testing,%Frontiers: Constrained surrogate testing of nonlinear dependence in bivariate time series
% barnes2024phase,%Chaos: Apply and describe constrained surrogates for investigating coherence
% kato2024dynamic,%PRE: Use IAAFT constrained surrogates for investigating combustion time series
%
li2025mapping,%Nat. Comm.: Constrained surrogates based on random rotation to test significance of observed correspondence between tract reachability and cortical geometry 
%
% schreiber2000surrogate,%Physica D: Review paper on surrogates
%
% zhang2025brain,%PNAS Nexus: Constrained network and time series surrogates for comparing different ANN training strategies
ma2025predicting%Commun. Phys.: Uses classical constrained surrogates to train ML for critical transition prediction
% christensen2015universality%J. Royal Soc. Interface: Uses a shuffling method to apply a null model
}. By reducing sensitivity to model misspecification, constrained surrogates can---especially at low sample length---provide higher accuracy and power in significance tests, avoid biasing estimates of arbitrary quantities, and enable the principled use of arbitrary (test) statistics~\cite{theiler1996constrained,moore2022nonparametric}.

A central question in statistical physics and complex systems is whether observations are consistent with a given parametric probability distribution $p(x)$, and how such distributions can be used to characterize and model the underlying system~\cite{newman2005power,broido2019scale,corral2020distinct,serafino2021true,altmann2025statistical,corral2019power}. In particular, fat-tailed distributions are often taken as evidence for underlying critical behavior~\cite{bak1987self,sornette2006critical,frette1996avalanche,corral2008scaling,nicoletti2020scaling,nicoletti2023emergence,wang2025atmospheric}. While constrained surrogates have been developed for discrete powerlaw models~\cite{moore2022nonparametric}, many other parametric distributions of comparable importance lack constrained surrogates which would allow theoretically supported assessment and application to empirical data. These include exponential~\cite{gutenberg1944frequency,altmann2012origin}, Gaussian~\cite{landau1980statistical,berg2025random}, and lognormal~\cite{mitzenmacher2004brief,pearson2018log} distributions, which arise from distinct generative mechanisms and are widely used to model diverse empirical systems.

In this paper, we introduce an approach for generating constrained surrogates representing arbitrary families of continuous distributions. 
Applying this approach, we obtain constrained-surrogate algorithms for the most central distributions in complex systems and statistical physics: continuous powerlaw, lognormal, exponential and truncated Gaussian distributions. We then show how the theoretical license to employ arbitrary statistics can enhance both hypothesis testing and statistical estimation in practice.  
We illustrate these advantages across several empirical systems. In particular, for data on the magnitude of Australian bushfires, constrained surrogates allow us to distinguish agreement in the bulk of the distribution from systematic overestimation of the probability of extreme events. Applications to word usage in texts and brain connectomes further illustrate how constrained surrogates enable tests based on physically meaningful quantities that reveal both the presence and the nature of discrepancies invisible to standard goodness-of-fit tests. Across all cases, these improvements are achieved without specifying or restricting model parameters.

\begin{figure*}[ht]
    \centering
    \includegraphics[width=\textwidth]{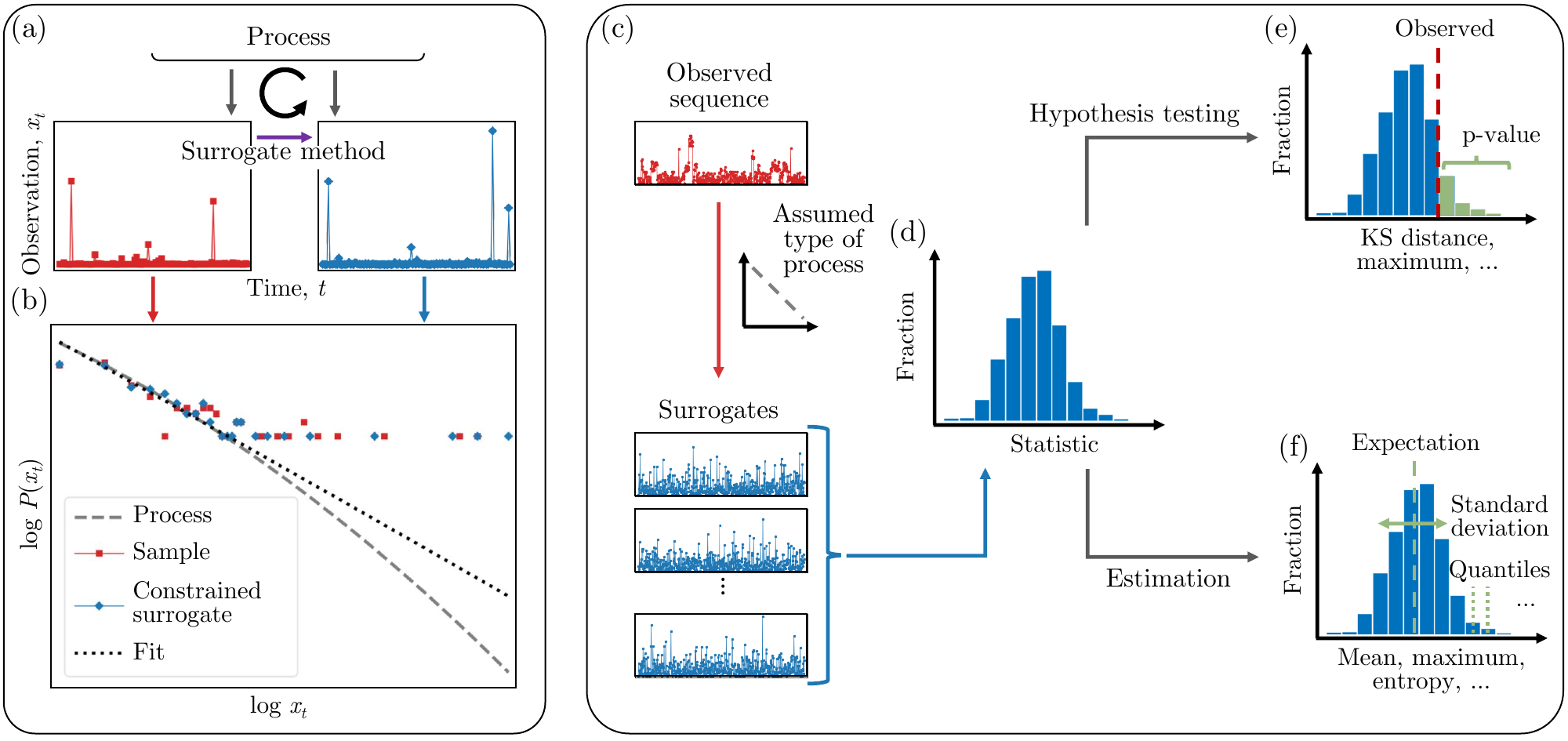}
    \caption{Surrogates in theory and practice. (a-b) Theory. (a) Constrained surrogates provide commutativity (circular arrow) in the sense that generating from a probabilistic process and then applying a constrained surrogate algorithm which properly represents that process is equivalent to generating directly from the original process: the two procedures yield the same marginal distribution. (b) A constrained surrogate preserves the likelihood and hence the maximum likelihood fit of an input sequence. Ideally, it is chosen uniformly at random from among all sequences which exhibit this likelihood function (and maximum likelihood fit). (c-e) Practice. (c) An assumed type of process (e.g., powerlaw, lognormal or exponential) applied to an observed sequence gives rise to a ensemble of surrogates. (d) A statistic of interest is calculated separately from each generated surrogate sequence. (e) Hypothesis testing is performed by comparing the value of the statistic calculated from the original sequence with the distribution estimated from surrogates and can lead to a rejection of the assumed process. (f) The distribution of statistics under surrogates is used to estimate the properties of the observed system, assuming it follows the hypothesized type of process.}
    \label{fig.illustration1}
\end{figure*}

\section{Typical and constrained surrogates}
 
A surrogate is a version of an observed dataset that was artificially generated under a hypothesized model.
While the process underlying empirical observations is often \emph{a priori} unknown, properly generated surrogates represent a hypothesis or model unambiguously, and this concreteness is key to their utility. Surrogates inform choice of model because they allow the observed dataset to be compared with datasets known to have arisen under a hypothesis. Once a model has been chosen, surrogates allow estimation of properties of the observed system based on the observed dataset, sometimes without the need to estimate parameters. We illustrate the role of surrogates in scientific research in Fig.~\ref{fig.illustration1}.

The \emph{typical} approach for obtaining surrogates involves estimating the parameters $\xi$ of the hypothesized model from the observed time series, e.g., through maximum likelihood, and then generating new time series as realizations of the model with fitted parameters $\xi = \hat{\xi}$. A prominent example of this approach is the test of powerlaw distribution, in which the parameter corresponds to the powerlaw exponent $\xi=\gamma$ and the model is sampling independent and identically distributed (i.i.d.) from the distribution with the estimated exponent~\cite{clauset2007frequency}. This method is convenient and intuitive, but has the limitation that it does not represent the hypothesized process in general (e.g., a powerlaw process with unknown exponent $\gamma$) but only the specific process with the parameters of best fit (e.g., powerlaw process with exponent $\gamma = \hat{\gamma}$). The ubiquity of correlations in real-world complex systems and the difficulty of accurately estimating parameters from limited samples can limit the accuracy and power of hypothesis tests based on typical surrogates and lead to bias in statistical estimates. The popular bootstrapping method (sampling with replacement) can be seen as an example of a typical surrogate method, built on the hypothesis that observations arose i.i.d. under some unknown distribution $p(x)$. Bootstrapping treats the empirical distribution of the observed data---the maximum likelihood estimate of $p(x)$---as the true distribution and generates surrogates by resampling from it.

The \emph{constrained} surrogate approach was originally described as an artificial dataset which exhibits the same parameters of best fit under the considered model as the observed data from which it arose~\cite{theiler1996constrained}. This definition has been refined to a dataset generated from the hypothesized process while conditioning on a sufficient statistic, which renders irrelevant the parameters of the underlying process~\cite{van1998testing}. We will employ a restriction of this definition which in many cases of interest is equivalent. Denoting by $\mathcal{L}_{{\bf x}}$ the likelihood of a sequence ${\bf x}$, a constrained surrogate for an input sequence $\xin$ is chosen uniformly at random from a set of sequences $\mathcal{S}\left(\xin\right)$ satisfying the following three conditions~\cite{moore2022nonparametric}:
\begin{itemize}
    \item[(1)] each sequence in the set has the same likelihood under the hypothesized process, i.e., $\forall {\bf x}, {\bf x}' \in \mathcal{S}\left(\xin\right), \mathcal{L}_{{\bf x}} = \mathcal{L}_{{\bf x}'}$; 
    \item[(2)] the set contains the observed sequence, i.e., $\xin \in \mathcal{S}\left(\xin\right)$; and 
    \item[(3)] the same set would be derived from any sequence which it contains, i.e., $\forall {\bf x} \in \mathcal{S}\left(\xin\right), \mathcal{S}\left({\bf x}\right) = \mathcal{S}\left(\xin\right)$.
    \end{itemize}
These conditions ensure that samples generated under the hypothesized process and their constrained surrogates have the same marginal distribution (see Fig.~\ref{fig.illustration1}a, circular arrow).

In addition to being parameter agnostic, other statistical advantages of constrained surrogates include unbiased estimation of \emph{any} test statistic, reduced variance, and exact hypothesis tests for arbitrary test statistics or sample lengths~\cite{theiler1996constrained,theiler1997using,besag2013exact,moore2022nonparametric}.

Here we are particularly interested in constrained surrogates representing the hypothesis that observations arose (i.i.d.) under some family of distributions $\left\{p_\xi(x)\right\}$. An approach allowing us to remain completely agnostic to the family of distributions would be to generate samples without replacement, i.e., shuffling. The above definition of a constrained surrogate is satisfied in this illustrative case because: (0) the sequence is chosen uniformly at random from among all sequences reachable via shuffling; (1) the likelihood of a sequence of i.i.d. observations does not depend on the order in which its elements appear; (2) the originally observed sequence can be generated by shuffling; and 
(3) the set of sequences reachable by randomizing the order of a previously shuffled sequence is the same as the set reachable from the sequence originally observed. Shuffling avoids assumptions about $p(x)$, which entails strong restrictions on realizable sequences. Specifically, shuffling preserves all order-insensitive properties of the original time series, including, e.g., a sample's minimum, maximum, median, quantiles, mean, and variance. Therefore, while shuffling is useful to test the i.i.d. hypothesis, it is of limited interest in other applications. In particular: as shuffling does not generate any sample $x_i$ that has not been previously observed, it cannot be used to estimate the probability of unobserved future events; and because shuffling simultaneously represents all families of distributions, it cannot be used to reach conclusions specific to a particular distributional family. In the next section, we introduce alternative constrained surrogate approaches that utilize assumptions about the functional form of $p(x)$ (though not its parameters) to 
overcome these limitations of shuffling as well as the bias and additional parametric assumptions of the typical surrogate approach. Thereby we achieve a balance between: (i) shuffling, which avoids bias and assumptions about distribution or parameters, but restricts realizable sequences such that no new observation can be generated; and
(ii) typical surrogates, which can provide a nonzero chance of realizing any sequence, but restrict both distribution and parameters, and incur bias. 

\section{Generating the surrogates}\label{sec.gen}

\subsection{General approach}

\begin{figure}[ht]
    \centering
    \includegraphics[width=\columnwidth]{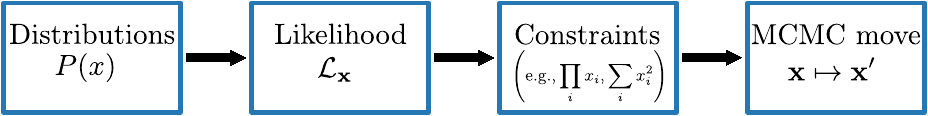}
    \caption{The steps involved in the generation of constrained surrogates.}
    \label{fig.steps}
\end{figure}

\paragraph*{Goal:} Given an input sequence $\xin$, our goal is to generate surrogate sequences ${\bf x}$ that preserve the likelihood $\mathcal{L}_{\bf x} (\xi)$ that the dataset was generated independently from a distribution within a family $\{p_\xi (x)\}$ spanning all parameter values $\xi$. More precisely, we wish to sample with uniform probability all sequences ${\bf x}=x_1,x_2, \ldots, x_N$ satisfying
\begin{equation}\label{eq.likelihood}
\mathcal{L}_{\bf x} (\xi) := \prod_{i=0}^N p_\xi(x_i) = \mathcal{L}_{\xin} (\xi),
\end{equation}
independently of $\xi$.

\paragraph*{Approach:} In Fig.~\ref{fig.steps} we illustrate our general idea to achieve the goal stated above. First, we identify distribution-specific constraints that preserve the likelihood in Eq.~(\ref{eq.likelihood}). Then we design a Markov Chain Monte Carlo (MCMC) approach enforcing these constraints via simple local moves. More specifically, at each iteration $r \mapsto r+1$, we randomize a $k$-subset $x_1,x_2, \ldots, x_k$ while satisfying the identified constraints. The symmetry of these transitions, and the possibility of reaching any sequence in $\mathcal{S}\left(\xin\right)$ (ergodicity), guarantees asymptotic ($r\rightarrow\infty$) uniform sampling from the space of sequences which satisfy the required constraints~\cite{robert2004monte}. 

{
\renewcommand{\arraystretch}{2.5}
\begin{table*}[ht]
    \caption{The likelihood function of important probabilistic models $p(x)$ is specified by one or more properties. Surrogates representing the model can be generated by randomizing the sequence while constraining these properties. The stated likelihood assumes the observed sequence sequence $x_1,x_2, \ldots, x_N$ is i.i.d. In cases for which the constrained properties are ``$x_1, \ldots, x_N$'', the only possible constrained surrogates are reachable by shuffling. In each case the constant $C$ is defined such that the total probability is unity.
    }
    \label{tab.likelihoods}
    \centering
    {
    \footnotesize
    \begin{ruledtabular}
    \begin{tabular}{c c c c c}
      Name & Probability of $x$& Support & Likelihood & \makecell[b]{Constrained\\ properties} \\ \hline
         Powerlaw &  $C x^{-\gamma}$ & $\left[x_{\min}, \infty\right)$ & $C^N \left( \prod_i x_i \right)^{-\gamma}$& $\prod_i x_i$\\
         Lognormal & $C \exp{\left(-\frac{1}{2}\left(\frac{\log x - \mu}{\sigma}\right)^2\right)}/x$ & $\left[x_{\min}, \infty\right)$ & 
         \makecell{$C^N \left( \prod_i x_i\right)^{-1}\exp\left(-\frac{1}{2 \sigma^2} \left( \sum_i \left(\log x_i\right)^2\right.\right.$\qquad\ \\
         $\qquad\qquad\qquad\qquad\ \ \left.\left.- 2 \mu \log \prod_i x_i + N \mu^2\right) \right)$}
         & $\prod_i x_i$, $\sum_i \left(\log x_i\right)^2$
         \\
         Exponential & $C \exp \left(-\lambda x \right)$ & $\left[x_{\min}, \infty\right)$ & $C^N \exp\left( -\lambda \sum_i x_i \right)$ & $\sum_i x_i$\\
         Gaussian & $C \exp{\left(-\frac{1}{2}\left(\frac{x - \mu}{\sigma}\right)^2\right)}$ & $\left[x_{\min}, \infty\right)$ & $C^N \exp\left(-\frac{1}{2 \sigma^2} \left( \sum_i {x_i}^2 - 2 \mu \sum_i x_i + N \mu^2\right) \right)$ & $\sum_i x_i$, $\sum_i {x_i}^2$\\
         Uniform & $\left(b - x_{\min}\right)^{-1}$ & $\left[x_{\min}, b\right]$ & $\left(b - x_{\min}\right)^{-N}$ & $\max\limits_{i \in \{1, \ldots, N\}} x_i$\\
         Powerlaw, exp. cut-off & $C x^{-\gamma} e^{-\lambda x}$ & $\left[x_{\min}, \infty\right)$ & $C^N \left( \prod_i x_i \right)^{-\gamma} \exp\left( -\lambda \sum_i x_i \right)$ & $\sum_i x_i$, $\prod_i x_i$\\
         Weibull, unknown $k$ & \multirow{2}{*}{$\frac{k}{\lambda} \left(\frac{x}{\lambda}\right)^{k-1} \exp\left(-(x/\lambda)^k\right)$} & \multirow{2}{*}{$\left[x_{\min}, \infty\right)$} & \multirow{2}{*}{$(\lambda^{-1} k)^N\left(\lambda^{-N} \prod_i x_i\right)^{k-1}\exp\left(-\lambda^{-k}\sum_i {x_i}^k\right)$} & $x_1, \ldots, x_N$\\
         Weibull, known $k \notin \{0,1\}$ & & & & $\prod_i x_i, \sum_i {x_i}^k$\\
         Shifted powerlaw, unknown $\delta$ & \multirow{2}{*}{$C (x + \delta)^{-\gamma}$} & \multirow{2}{*}{$\left[x_{\min}, \infty\right)$} & \multirow{2}{*}{$C^N \left[ \prod_i \left( x_i + \delta \right) \right]^{-\gamma}$} & $x_1, \ldots, x_N$\\
         Shifted powerlaw, known $\delta$ & & & & $\prod_i \left( x_i + \delta \right)$
    \end{tabular}
    \end{ruledtabular}
    }
\end{table*}
}

\begin{figure*}
    \centering
    \includegraphics[width=\textwidth]{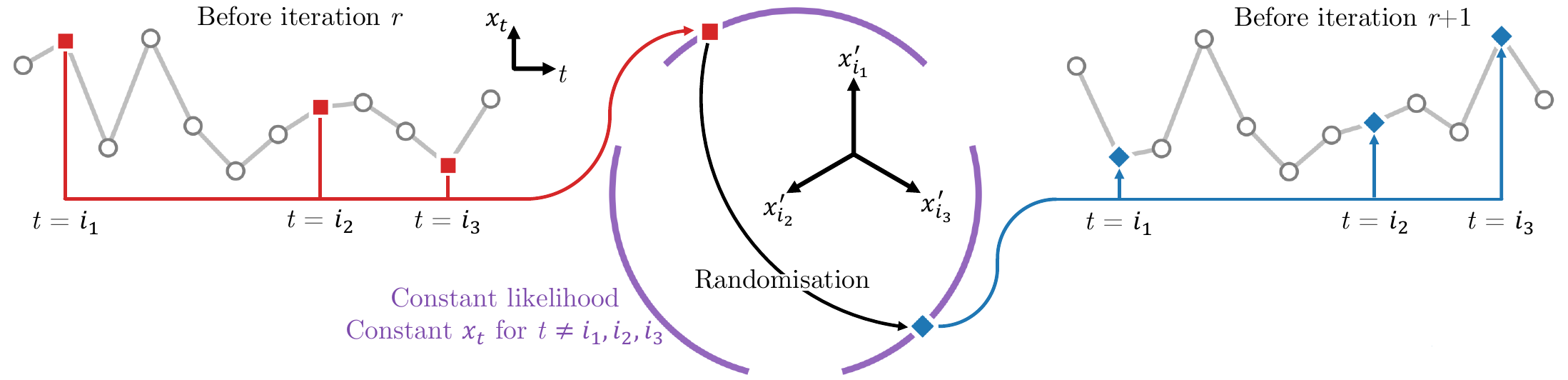}
    \caption{Illustration of the proposed algorithm for generating constrained likelihood surrogates. The figure concentrates on one MCMC step for the case of a truncated ($x\geq x_{\min}$) Gaussian distribution. In this case,  the likelihood is specified by $K = 2$ continuous properties which we must preserve to produce constrained surrogates (the same general approach applies to other distributions). In each iteration $r$ we randomly choose $K + 1$ elements $x_{i_1}, \ldots, x_{i_{K+1}}$ (red squares) from the sequence. These correspond (red line) to a point (red square) on the curve (purple arcs) giving the collection of points in $x_{i_1}'\cdots x_{i_{K+1}}'$-space 
    (black coordinate axes) which can replace $x_{i_1}, \ldots, x_{i_{K+1}}$ without altering the likelihood. 
    The next location (blue diamond) is selected uniformly at random (black arrow) and the sequence is updated according to $x_{i_1} = x_{i_1}', \ldots, x_{i_{K+1}} = x_{i_{K+1}}'$ (blue arrows).}
    \label{fig.metropolis}
\end{figure*}

\subsection{From distributions to constraints}\label{sec:distributions}

The key observation underlying our approach is that the likelihood in Eq.~(\ref{eq.likelihood}) can be kept constant by imposing a few specific constraints on the sequence $\{x_i\}$. For instance, constraining the likelihood of observations under the assumption of an i.i.d. truncated Gaussian distribution corresponds to fixing the sums $\sum_i x_i$ ($\propto$ first moment) and $\sum_i {x_i}^2$ ($\propto$ second moment). Our MCMC implementations discussed below succeed in fixing the following constraints:
\begin{enumerate}
    \item Thresholds: $x\geq x_{\min}$ or $x\leq x_{\max}$;
    \item Geometric mean $\left(\prod_{i=1}^N x_i\right)^{1/N}$ or product $\prod_{i=1}^N x_i$;
    \item Mean $\frac{1}{N}\sum_{i=1}^N x_i$ or sum $\sum_{i=1}^N x_i$;
    \item Second moment $\sum_{i=1}^N {x_i}^2$.
\end{enumerate}
We also combine two or more of these constraints and consider transformation of variables $y_i=f(x)$, e.g., $y_i= \log x_i$. 
Table~\ref{tab.likelihoods} shows how these allow us to tackle some of the most common distributions used in statistical physics and time-series analysis.

\subsection{From constraints to MCMC methods}\label{sec:mcmc}

We now construct MCMC methods that sample from sequences {\bf x} that satisfy the constraints of Sec.~\ref{sec:distributions},
targeting uniform (i.e, constant probability) sampling of the $(N-K)$-dimensional manifold generically formed by imposing $K$ constraints over $N$ data points. For instance, in the Gaussian case, this corresponds to sampling uniformly from the set of sequences that have the same first and second moment as the input sequence. We achieve this with localized transitions where in each step $r \mapsto r+1$ we propose a move ${\bf x} \mapsto {\bf x}'$. We use symmetric transitions---$P({\bf x} \rightarrow {\bf x'}) = P({\bf x'} \rightarrow {\bf x})$---and ensure that all entries of ${\bf x}$ can be modified, leading to  a Markov chain that (for $r\rightarrow \infty$) ergodically explores the constraint manifold with uniform probability~\cite{robert2004monte}.
 
In each iteration we randomize $K + 1$ of the $N$ values of ${\bf x}$,
\begin{equation} {\bf x}_{K+1}:=x_{i_1}, \ldots, x_{i_{K+1}} \mapsto x'_{i_1}, \ldots, x'_{i_{K+1}}:= {\bf x}'_{K+1},
\end{equation}
where the integers $i_1, \ldots, i_{K+1} \in [1, N]$ are chosen uniformly at random without replacement. Our aim is for the move to satisfy all the global constraints so that our exploration is restricted to the allowable sequences. We thus compute the constraints in $x_{i_1}, \ldots, x_{i_{K+1}}$ and impose the same constraints into $x'_{i_1}, \ldots, x'_{i_{K+1}}$.
In the Gaussian example, $K=2$ and we will therefore choose $3$ random points from ${\bf x}$ to modify without altering their sum and sum of squares. Generically, the choice about how to modify them is a 1-dimensional problem because the $K$ constraints will be satisfied in some subset of a curve in $\mathbb{R}^{K+1}$ parametrized by $\gamma(s)$, for $s \in [a,b]$. The choice of the move (new values of the points) is thus reduced to the random choice of $s \in [a,b]$ that determines ${\bf x}'$. 

The distribution $P(s)$ of $s \in [a,b]$ implies the probability with which sequences ${\bf x}$ are sampled. To guarantee symmetric transition probabilities, we choose destinations uniformly at random along a 1-manifold (made up of one or more curves) of points which preserve the $K$ constrained properties. To see how the desired uniform sampling can be achieved, we transform to coordinates $\bbeta, s$, where $\bbeta$ is the $K$-vector listing the $K$ constrained properties. 
In these coordinates, the probability of choosing a point ${\bf x}_{K+1}$ is
\begin{equation}\label{eq.P_x_{K+1}}\notag
P \left( {\bf x}_{K+1} \right) = P \left( \bbeta \right) P(s)
\left\lvert \det \pp{\bbeta, s}{{\bf x}_{K+1}} \right\rvert, 
\end{equation}
where $\pp{\bbeta, s}{{\bf x}_{K+1}}$ is the Jacobian matrix of the transformation from coordinates $\mathbf{x}_{K+1}$ to coordinates $\bbeta,s$. Uniform sampling---i.e., constant $P \left( {\bf x}_{K+1} \right)$---is therefore equivalent to a product $P(s)
\left\lvert \det \pp{\bbeta, s}{{\bf x}_{K+1}} \right\rvert$ which is independent of $s$, or to
\begin{equation}
    P(s) = C \left\lvert \det \pp{\bbeta, s}{{\bf x}_{K+1}} \right\rvert^{-1}, \label{eq.prob_dens_t}
\end{equation}
where $C$ is a normalization constant.

{
\begin{ruledtabular}
\begin{table*}[ht]
    \caption{Key ingredients for constrained surrogates representing five important families of distributions. In each iteration of the algorithm to generate a constrained surrogate representing a given distribution, we randomly choose $K + 1$ elements $x_{i_1}, \ldots, x_{i_{K+1}}$ and from these determine the values of the $K$ properties $\beta_1, \ldots, \beta_K$ to be constrained. We then randomize the values of the elements $x_{i_1}, \ldots, x_{i_{K+1}}$ by picking a point along the curve $\gamma: s \mapsto \gamma(s)$ by choosing $s$ with probability $P(s)$ from the specified domain. $R$ is a fixed $3 \times 3$ rotation matrix (see Appendix~\ref{app.specific}), $\rho \left( \bbeta \right) = \frac{1}{\sqrt{3}} \sqrt{3 \beta_2 - {\beta_1}^2}$, and the exponential function $\exp$ is applied element-wise.
    }
    \label{tab.constraints}
    \centering
    {\small
    \begin{tabular}{c c c c c}
      Name  & Constrained properties $\bbeta$ & Curve $\gamma: \mathbb{R} \mapsto \mathbb{R}^{K+1}, s \mapsto \left( x_{i_1}, \ldots, x_{i_{K+1}}\right)$ & Probability $P(s)$ & Domain of $s$\\ \hline
     Exponential & $\beta_1 = x_{i_1} + x_{i_2}$ & $\left(s, \beta_1 - s\right)$ & ${\beta_1}^{-1}$ & $[0, \beta_1]$\\
     Gaussian & $\beta_1 = \sum\limits_{j=1}^3 x_{i_j}$, $\beta_2 = \sum\limits_{j=1}^3 {x_{i_j}}^2$ & $r(\bbeta) R \cdot \left( \cos t, \sin t, 0\right)^T + \left(1, 1, 1\right)^T \beta_1/3$ & $1/2 \pi$ & $[0, 2\pi)$\\
     \multirow{ 2}{*}{Powerlaw} & \multirow{ 2}{*}{$\beta_1 = x_{i_1} x_{i_2}$}
     & $\left( s, \beta_1/s \right)$ & $\left[ s \log (\beta_1)\right]^{-1}$ & $[1, \beta_1]$\\
     & & $\left( \exp s, \beta_1/\exp s \right)$ & $\log(\beta_1)^{-1}$ & $[0, \log(\beta_1)]$\\
     Lognormal & $\beta_1 = \prod\limits_{j=1}^3 x_{i_j}$, $\beta_2 = \sum\limits_{j=1}^3 \left(\log x_{i_j}\right)^2$ & $\makecell{\exp \left[\rho(\bbeta) R \cdot \left( \cos s, \sin s, 0\right)^T\right.\\ \qquad\ \ \ \left. + \left(1, 1, 1\right)^T\left(\log \beta_1\right)/3\right]}$ & $1/2 \pi$ & $[0, 2\pi)$\\
     Uniform\footnote{To generate constrained surrogates representing the uniform distribution we must preserve the maximum value but do not need other components of our framework (see Appendix~\ref{app.specific}).} & $\beta_1 = \max\limits_{i \in \{1,\ldots,N\}} x_{i}$ & - & - & -
    \end{tabular}
    }
\end{table*}
\end{ruledtabular}
}

An illustration of one MCMC step in the case of truncated Gaussians is shown in Fig.~\ref{fig.metropolis}. All the cases in which we implement this general framework are summarized in Table~\ref{tab.constraints} (for their derivations, see Appendix~\ref{app.specific}). Constrained uniform surrogates can be generated without iteration. An important practical point is to determine the number of MCMC transitions to be used in the sampling. In our case, we can show that (see Appendix~\ref{app.transitions})
\begin{equation}\label{eq.rtilde}
    \tilde{r} = \frac{N}{2}\left(\log_2 N - \log_2 \epsilon \right)
\end{equation} 
transitions are sufficient to achieve a relative error in the expected value (of either a sequence of length $N$ or its log-transform) of no more than $\epsilon$ . In our simulations, unless stated otherwise, we use $2 \tilde{r}$ (with $\epsilon = 2^{-10}$) transitions between samples surrogate samples and multiple samples are
%For further randomization, unless stated otherwise, we generate a constrained surrogate from each observed sequence and generate each subsequent constrained surrogate from the previous constrained surrogate.
%Unless stated otherwise, for a given observed sequence, each constrained surrogate other than the first is 
generated starting from the previous one. %constrained surrogate.}

\begin{table*}[!htbp]
	\caption{
		The probability density functions used to generate our synthetic data with lower cutoff $x_{\min}=1$. In each case the constant $C$ is defined such that the total probability is unity.
	}
	\label{tab.distributions}%
    \begin{ruledtabular}
		\begin{tabular}{c c c c}
			Name & Probability & Support & Parameter(s)\\
			\hline
			Exponential
			& $p(x) = C x^{-\lambda}$ & $\left[x_{\min}, \infty\right)$ & $\lambda = 1$\\
			Gaussian (truncated)
			& $p(x) = C \frac{1}{\sigma \sqrt{2 \pi}} \exp \left[ -\left( \frac{x - \mu}{\sqrt{2} \sigma}\right)^2 \right]$ & $\left[x_{\min}, \infty\right)$ & $\mu = -1$,
			$\sigma = 1$\\
			Powerlaw
			& $p(x) = C x^{-\gamma}$ & $\left[x_{\min}, \infty\right)$ & $\gamma = 2.5$\\
			Lognormal (truncated)
			& $p(x) = C \frac{1}{x \sigma \sqrt{2 \pi}} \exp \left[ -\left( \frac{\log x - \mu}{\sqrt{2} \sigma}\right)^2 \right]$ & $\left[x_{\min}, \infty\right)$ & $\mu = -1$,
			$\sigma = 1$\\
			Uniform
			& $p(x) = C$ & $\left[x_{\min}, b\right)$ & $b = 9$\\
		\end{tabular}
    \end{ruledtabular}
\end{table*}

\section{Performance}\label{sec.performance}

In this section we use synthetic data to illustrate the properties of constrained surrogates and explore their suitability for different tasks. Using synthetic data of the types listed in Table~\ref{tab.distributions}, we consider surrogates designed to represent the hypothesis that data arose i.i.d. under five different families of distribution $\left\{p_\xi(x)\right\}$: powerlaw, truncated lognormal, exponential, truncated Gaussian, and uniform. The surrogates we apply include constrained surrogates, as defined in Sec.~\ref{sec.gen}, and typical surrogates, generated by first determining the parameters $\hat{\xi}$ that maximize the likelihood of the hypothesized model and then generating i.i.d. sequences from the distribution $p_{\hat{\xi}}(x)$ with parameter $\hat{\xi}$.

\subsection{Hypothesis testing}

\begin{figure*}
    \centering\includegraphics[width=\textwidth]{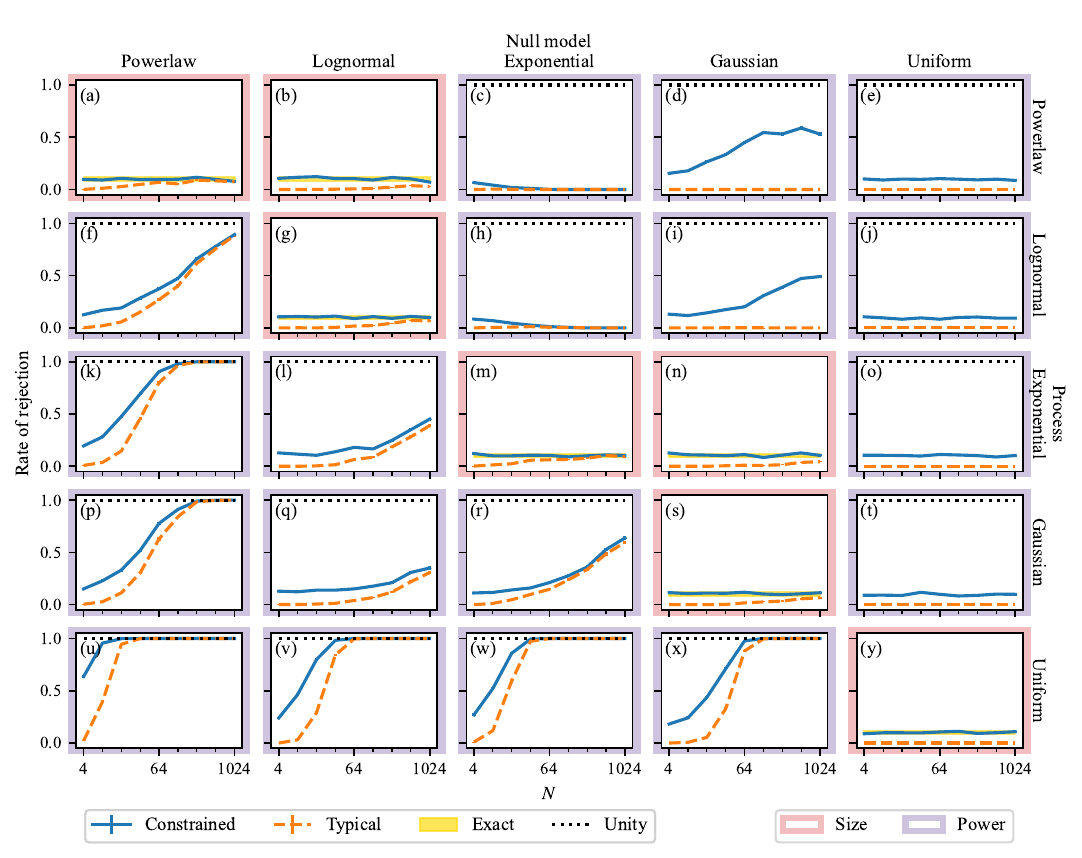}
    \caption{Constrained surrogates provide exact hypothesis tests for arbitrary discriminating statistics and can increase power to reject false hypotheses. The power (size) of the test of null hypothesis was estimated from $1,000$ tests, each using $N$ i.i.d. samples and 19 surrogates. Results are shown for different combinations of underlying distribution (from top to bottom) and null hypothesis (from left to right): powerlaw; lognormal;  exponential; truncated Gaussian; and uniform distribution. 
    Tests are lower-directed, have nominal size 10\%, and use as the sample maximum as discriminating statistic. Panels showing tests for which the null hypothesis is correct (i.e., for which the rate of rejection is the size of the test), comprising all panels on the diagonal and two off-diagonal panels, have red borders. These six panels feature a gold band which spans the 90\% confidence interval for a Bernoulli process, but the band is thin and is often obscured by the blue line showing the size achieved by constrained surrogates. Panels showing tests for which the null hypothesis is incorrect (i.e., for which the rate of rejection is the test's power to reject an incorrect hypothesis), comprising the remaining panels, have purple borders and feature a dotted black line corresponding to the ideal (100\%) rejection rate.
    Error bars correspond to standard error but are often indistinguishable from the line width.}
    \label{fig.hyp-max-left-tail}
\end{figure*}

First we consider using surrogates to test the validity of a hypothesized null model. As usual in hypothesis testing, this null hypothesis is rejected if the \emph{p-value}---defined as the probability under the null hypothesis of observing a value of a discriminating statistic at least as extreme as the value observed in the input sequence $\xin$---falls below a predetermined threshold parameter termed the \emph{nominal size} (frequently $0.05$ or $0.1$). Surrogates enable computationally efficient hypothesis testing because they allow this probability to be estimated by evaluating the discriminating statistic across an ensemble of surrogates $\{{\bf x}_n\}$.  
The standard method for assessing goodness-of-fit involves hypothesis tests that use as discriminating statistic the Kolmogorov-Smirnov (KS) distance, computed relative to the maximum likelihood parameters of the portion of the dataset above the lower cut-off~\cite{clauset2009power}. In this section, to illustrate the capacity of constrained surrogates to provide theoretical support for arbitrary discriminating statistics and also because of its physical relevance (e.g., for investigating extreme events), we instead employ as a test statistic the sample maximum, a quantity of interest in the study of extreme events. 
We focus on lower-directed hypothesis tests and quantify the performance of surrogate methods via two key metrics:
\begin{itemize}
    \item The \emph{size} of a test is the rate of incorrect rejection of the null hypothesis, i.e., rejection of the null hypothesis when it is actually true. 
    Ideally a test is \emph{exact} or \emph{accurate}, which occurs when its size equals the predetermined nominal size. 
    \item The \emph{power} of a test is the rate of correct rejection of the null hypothesis, i.e., rejection of the null hypothesis when it is incorrect. This rate depends not only on surrogate method and discriminating statistic, but also on the true generating process which underlies the input sequence.
\end{itemize}

In Fig.~\ref{fig.hyp-max-left-tail} we consider both accuracy and power. To assess size (in panels with red borders) we generate $\xin$ as an i.i.d. sample from a given family of distributions and test whether different types of surrogates reject the hypothesis that the sample arose under this distribution or a generalization (powerlaw and exponential distributions are a limiting case of truncated lognormal and Gaussian distributions). We compare how well constrained and typical surrogates allow us to match the hypothesis test's true size with a chosen nominal size. We see that typical surrogates consistently undershoot nominal size for small sample lengths $N$ and that, for sequences arising under lognormal, Gaussian or uniform processes, this deviation remains evident even for the largest sample lengths considered. This underestimation can be interpreted as advantageous because it reduces the probability of incorrectly rejecting a true exponential, Gaussian, or lognormal process, but also entails inaccuracy, because size should be a parameter which can be specified before testing~\cite{theiler1996constrained}. Furthermore, a mismatch between the true and nominal size is inherently disturbing, because it implies we are not correctly representing the null hypothesis. Turning to power (in panels with purple borders), we generate $\xin$ from a distribution and test whether different surrogate methods can reject the incorrect hypothesis that the data arose from a distinct probability functions. We see that constrained surrogates frequently provide higher power to reject an incorrect hypotheses. When the null model is uniform, this higher power equals the nominal size. This occurs because constrained surrogates preserve the maximum-likelihood estimate of the model parameter, which in this case is also the discriminating statistic. Other tests show different patterns: in upper-directed tests also using sample maximum or using KS distance, typical surrogates can offer higher power (see Supplemental Material (SM)~\cite{supplemental}, Fig.~\ref{fig.hyp-max-right-tail},\ref{fig.hyp-KS-dist}); and for two-tailed tests involving log-kurtosis or coefficient of variation the more powerful surrogate depends on both the true underlying process and the null hypothesis (see SM~\cite{supplemental}, Fig.~\ref{fig.hyp-coef-var},\ref{fig.hyp-skewness}). 
As a rule of thumb, we found that for common descriptive statistics, constrained (typical) surrogates tended to be better for lower-directed (upper-directed) tests (see SM~\cite{supplemental}, Fig.~\ref{fig.hyp-various-stats-left-tail},\ref{fig.hyp-various-stats-right-tail}). The generality of this rule is surprising given that a lower-directed test using a given test statistic is equivalent to an upper-directed test using the negation of that test statistic.

% \change{When generating each constrained independently from the original sequence, after $\tilde{r}$ or even $2 \tilde{r}$ transitions, statistical power has not yet plateaued (see Fig.~\ref{fig.vs-nt-hyp-max-left-tail}), which suggests that some of the statistical power associated with constrained surrogates is due to our choice to continue the sequence of transitions until the required number of surrogates is reached. It also suggests that further improvements to statistical power could be achieved by increasing the number of transitions used or the number of surrogates. However, this would also increase their computational cost, which grows with transition count and data length (see Figs.~\ref{fig.comp-cost-vs-N},\ref{fig.comp-cost-vs-num-trans}).}

Finally, we test whether our prescription of using $\tilde{r}$ MCMC transitions [see Eq.~(\ref{eq.rtilde})], which was obtained by bounding the relative error of a sequence, is also appropriate for hypothesis testing. To do so, we examine how size and power vary with the number of transitions used for fixed data length $N=64$. We found that size remains at or below nominal size and stabilizes to the nominal size by or just after $\tilde{r}$ transitions (see Fig.~\ref{fig.vs-nt-hyp-max-left-tail}). In almost all cases, power has also plateaued by $\tilde{r}$ transitions, although when the true process is powerlaw or lognormal, the power provided by constrained Gaussian surrogates can continue to increase beyond this number of transitions.

\subsection{Estimation}

\begin{figure*}
    \centering
    \includegraphics[width=\textwidth]{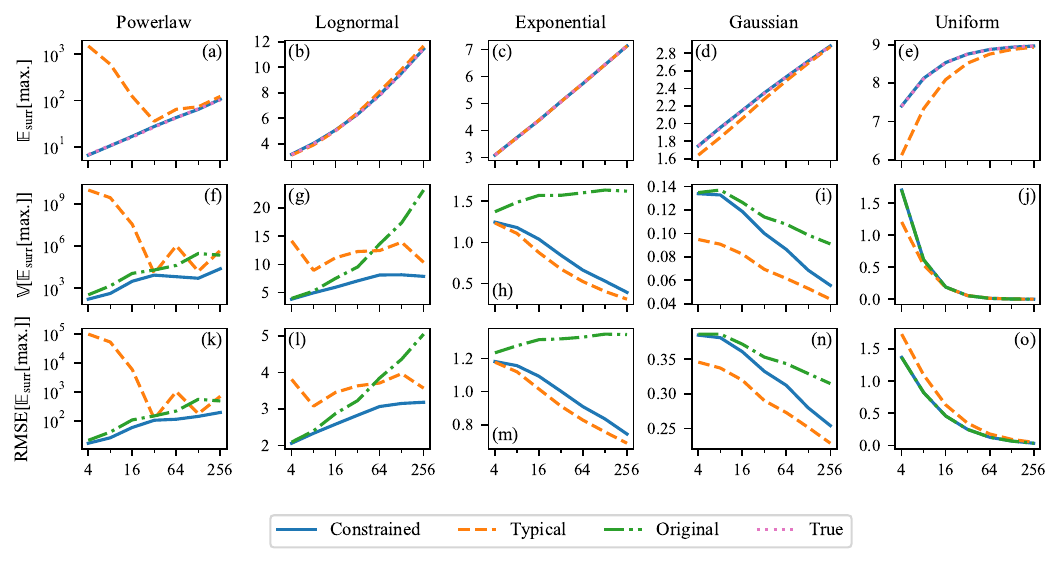}
    \caption{%
    Constrained surrogates do not bias sample statistics, and can reduce variance and error. Expectation (first row), variance $\mathbb{V}$ (second row), and root mean square error (RMSE, third row) of estimated expectation $\mathbb{E}_{\text{surr}}$ of sample maximum. The results are plotted against the length $N$ of the input sequence ${\bf x^{\rm{input}}}$ which is equal to the length of the surrogate sequences ${\bf x}$. $\mathbb{V}$ is calculated using $10^4$ independent input sequences ${\bf x^{\rm{input}}}$, each comprising $N$ values drawn i.i.d. from a (from left to right): powerlaw; lognormal; exponential; truncated Gaussian; and uniform distribution.
    $\mathbb{E}_{\text{surr}}$ is estimated using $100$ surrogates, each from the same ${\bf x^{\rm{input}}}$. The surrogates considered are either constrained (blue solid), typical (orange dashed), unchanged from the original observation (green dot-dashed), or generated independently from the true underlying process (pink dotted).
    }
    \label{fig.estimation}
\end{figure*}

Now we utilize surrogates in estimating properties of an underlying system, with a focus on its extreme events. 
For constrained surrogates, these estimates are based both on the observed time series and the model (family of distributions) assumed to describe the underlying process, but not the parameter value defining the precise process (e.g., assuming a powerlaw distribution but not the exponent $\gamma$). Specifically, we take as $\xin$ synthetically generated sequences from an underlying process, generate surrogates with the same size $N$ as $\xin$, and for each surrogate compute the sample maximum (i.e., the maximum of the $N$ values). 
The first row of Fig.~\ref{fig.estimation} reveals that expectation calculated using constrained surrogates coincides with expectation under the true process, implying that---as expected theoretically---constrained surrogates provide an unbiased estimate of the sample maximum. This contrasts with the typical approach, which leads to expectations that can either underestimate or overestimate, especially for lower sample length $N$. The second row shows that constrained surrogates also reduce the variance in this estimate relative to the original observed maximum and, for the heavy-tailed powerlaw and lognormal processes, also lead to lower variance than the biased estimate based on typical surrogates. Finally, from the third row we see that constrained surrogates consistently reduce error in estimates of sample expectation, and that in three of the five cases the improvement is greater than under typical surrogates. In some cases, particularly for low sample length $N$, typical surrogates can increase error in this estimate. 
Constrained surrogates also exhibit advantages for other statistics, especially when sample length $N$ is small (see SM~\cite{supplemental}, Fig.~\ref{fig.estimation-coef-var}).

\section{Applications}\label{sec.applications}

In this section we apply surrogate methods to estimation and hypothesis testing in data representing diverse types of empirical complex system.

\subsection{Bushfires} \label{ssec.fires}

\begin{figure*}
\includegraphics[width=\textwidth]{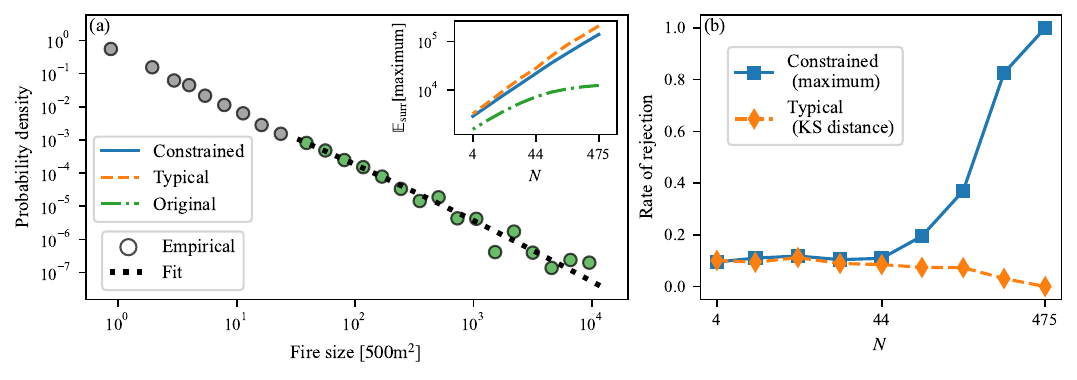}
    \caption{Constrained surrogates improve conclusions about the distribution of bushfire magnitudes. (a) Log-binned empirical distribution of fire magnitude (circles) and the maximum likelihood fit to a lognormal model (dotted black line) of values above the lower cutoff. Inset: Expectations of sample maximum estimated as the mean over $10^5$ independent samples of length $N$ downsampled (i.e., drawn i.i.d. without replacement) either from the original data or surrogates produced from the full dataset. (b) Rates of rejection estimated from 1,000 hypothesis tests, each using a sample of length $N$ from the observed dataset, 99 surrogates, and either the KS distance together with typical surrogates or maximum with constrained surrogates. Only values of the original data which exceed the fitted lower cutoff $x_{\min} = 32.5\ [500 \textrm{m}^2]$ are considered.
    }
    \label{fig.fires}
\end{figure*}

First, we consider a time series describing the magnitudes of bushfires that occurred in Australia from July 2019 until June 2020, representing one of Australia's worst recorded bushfire years. Bushfires were inferred from satellite imaging data as described in Ref.~\cite{nicoletti2023emergence} and their sizes, in units of 500 m\textsuperscript{2}, range from 1 to 12~446. 
The severity of bushfires in the considered period has been linked to a critical transition associated with powerlaw scaling of bushfire magnitude~\cite{nicoletti2023emergence}, in line with a long tradition of linking fires, natural disasters, extreme events, critical phenomena, and powerlaw distributions~\cite{bak1987self,malamud1998forest,sornette2006critical,corral2008scaling,altmann2025statistical}.

In Fig.~\ref{fig.fires} we investigate the hypothesis that the distribution of bushfire magnitudes is lognormal. This distribution contains powerlaws as a particular case so that we are also testing the most popular distribution. Visually, the log-binned empirical probability density convincingly follows a straight line over several orders of magnitude. The tails are also well described by the lognormal distribution up to the maximum observed bushfire magnitude, with the maximum likelihood fit deviating only imperceptibly from a straight line corresponding to a powerlaw distribution. In our quantitative analysis of the tail of the distribution we used the lower cut-off $x_{\min} = 32.5$ chosen via the widely used method of minimizing the KS distance relative to the maximum likelihood parameters~\cite{clauset2009power}. Repeating the analysis using $x_{\min}=16.5$ and $64.5$ produced the same qualitative results. Using the KS distance as discriminating statistic---as recommended in Ref.~\cite{clauset2009power}---to test the hypothesis that the dataset arose under a lognormal distribution, we observe rates of rejection of an i.i.d. lognormal hypothesis not much larger than the nominal size of the test (across a range of values of $N$, see Fig.~\ref{fig.fires}b). All this evidence---obtained following standard statistical recipes in the field~\cite{clauset2007frequency}---corroborates the connection between bushfires magnitudes and powerlaw/criticality models.

However, the conclusions above are in conflict with the strong deviations between the frequency of extremely large bushfires and the powerlaw/lognormal predictions. This is evident when comparing the observation and predictions of maximum fire magnitude (see inset of Fig.~\ref{fig.fires}a). Across the range of considered data lengths, the model predictions substantially exceed those based on subsampling the original dataset, although the use of constrained surrogates does slightly ameliorate these overestimates. 
Fortunately, constrained surrogates enable us to use arbitrary test statistics in hypothesis tests, and we therefore turn to a statistic of socioeconomic significance: the sample maximum. Using this statistic and constrained surrogates in lower-directed hypothesis tests, rates of rejection increase to unity as the sample length approaches that of the original dataset. For lower sample lengths, as expected from benchmarking, constrained surrogates can provide substantially higher rates of rejection than can typical surrogates (see SM~\cite{supplemental}, Fig.~\ref{fig.fires-supp}). Similar results are reached for the powerlaw model (see SM~\cite{supplemental}, Fig.~\ref{fig.fires-powerlaw-supp}). Altogether, constrained surrogates allowed us to reach the nuanced yet significant conclusions that, while powerlaws and lognormal distributions provide a good description---spanning four orders of magnitude---of the bulk of the Australian bushfire distribution, these models overestimate the probability of extremely large fires.

\subsection{Words}

\begin{figure}
\includegraphics[width=\columnwidth]{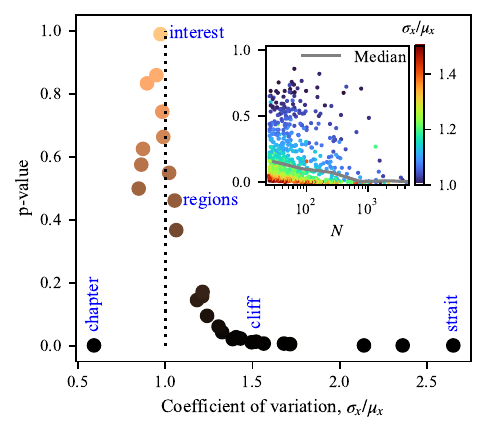}
    \caption{Constrained surrogates to test the distribution of words in texts. Reported is the outcome of the tests of an exponential distribution of waiting times $x$ between repetitions of the same word in the book \emph{The Voyage of the Beagle}, by Charles Darwin. The main panel shows the relation between the p-value of the test and the coefficient of variation $\sigma_x/\mu_x$ for words with $N=30$ waiting times, with five words highlighted and marker color indicating p-value.
    Inset: p-value versus $N$. Each marker corresponds to a word with $1 \leq \sigma_x/\mu_x < 1.5$, and its color indicates its corresponding coefficient of variation $\sigma_x/\mu_x$. The median p-value is shown as a function of $N$ (gray line). Hypothesis tests use 999 constrained surrogates and employ $\sigma_x/\mu_x$ as discriminating statistic. We consider the text's 1,000 most frequent words and a lower cutoff $x_{\min} = 9.5$, which approximates the typical length of a clause in written English~\cite{lyu2022marked}; repeating the analysis using $x_{\min} = 4.5$ and $x_{\min} = 19.5$ does not qualitatively change results.}
    \label{fig.words}
\end{figure}

Next, we investigate the distribution of repeated appearances of the same word in written English. We quantify this by counting the number $x$ of word tokens (waiting times) between word repetitions within a text. Deviations of the waiting times from an exponential distribution---expected from a Poissonian (i.e., memoryless) null hypothesis---have been used to characterize the burstiness and connected to different (syntactic and semantic) properties of the words~\cite{Ortuno2002keyword,altmann2009beyond,corral2009universal,altmann2025statistical}. 

In Fig.~\ref{fig.words} we consider using constrained surrogates to test the hypothesis that each waiting time sequence arose under an i.i.d.\ exponential process, employing as discriminating statistic the coefficient of variation $\sigma_x/\mu_x$ given by the ratio of the standard deviation to the mean. The coefficient of variation quantifies burstiness: for an i.i.d. exponential process and large sample sizes $N$, it approaches one. We compare our surrogate-based approach with the simpler strategy of directly inspecting the coefficient of variation.
For words with $N=30$ (main panel in Fig.~\ref{fig.words}), we see the direct connection between the p-value of our test, $\sigma_x/\mu_x$, and the role played by different words in the text:  function words have values with $\sigma_x/\mu_x \approx 1$ (e.g., ``interest''), topical words have $\sigma_x/\mu_x \gg 1$ are  (e.g., ``strait''), and the rare words with regular appearance have $\sigma_x/\mu_x \ll 1$  (e.g., the word ``chapter'': chapters of Darwin's text are of similar size and open with ``Chapter $n$''). In testing an i.i.d. exponential hypothesis, as expected for this low sample length, we see high p-values when coefficient of variation is close to unity. However, the observed p-values rapidly decrease as we move away from this value of coefficient of variation. Extending our scope to the words for which $1 \leq \sigma_x/\mu_x < 1.5$, we still observe p-values tending to decrease as coefficient of variation departs from a value of one. For a given coefficient of variation, p-values also decrease as $N$ increases, with fluctuations for the largest values of $N$, where data are scarce. Other ranges of $\sigma_x/\mu_x$ exhibit analogous patterns (see SM~\cite{supplemental}, Fig.~\ref{fig.words-p-vs-N}).
This reflects how, for a sufficiently long sample, even a small departure from a hypothesized family of distributions becomes statistically significant. Our constrained surrogate-based approach to testing for an exponential distribution thus accounts both for the burstiness (quantified by $\sigma_x/\mu_x$) and also the sample size $N$.

\subsection{Brain data}

\begin{figure*}
\includegraphics[width=0.98\linewidth]{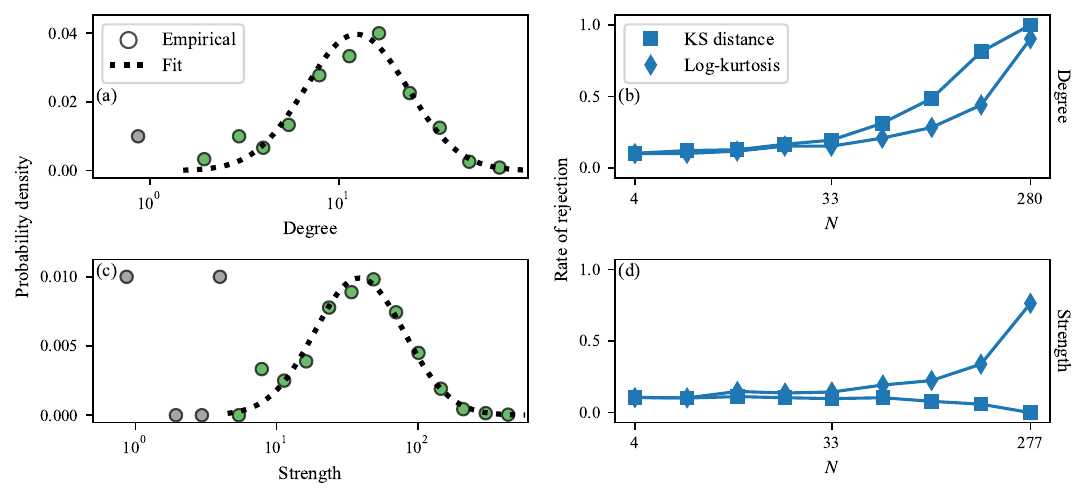}
    \caption{
    Constrained surrogates lead to improved characterization of neuron connectivity. 
    Conclusions about the validity of lognormal models depend on whether we focus on goodness-of-fit or fatness of tails. (a)-(b) Log-binned empirical distribution (circles) of (a) degree (number of other neurons with which a synaptic connection is shared) and (b) strength (number of synaptic connections with other neurons) in the neuron-resolution connectome of \emph{C. elegans.}, together with the maximum likelihood fit to a lognormal model (dotted black line). (c-d) Rates of rejection estimated from 1,000 hypothesis tests, each using a sample of length $N$ from the sequence of observed (c) degrees and (d) strengths above the lower cutoff, 99 constrained surrogates and either the KS distance (measuring goodness of fit) or log-kurtosis (measuring fatness of tails). Only values which equal or exceed the lower cutoffs $x_{\min} = 1.5$ (for degree) and $x_{\min} = 4.5$ (for strength) are considered.
    }
    \label{fig.brain}
\end{figure*}

Now, we turn our attention to neuron connectivity. Specifically, we examine the sequence listing each neuron's degree (number of other neurons with which a synaptic connection is shared) and strength (number of synaptic connections with other neurons), both of which have been described as lognormally distributed across all neurons in the brain of the nematode \emph{C. elegans}~\cite{piazza2025physical}. Of particular interest in debates about distributions in complex systems is the fatness of their tails, which can be assessed with a range of statistics~\cite{voit2005statistical}. In our case an especially pertinent measure is the excess kurtosis of the log transformed data (hereafter called ``log-kurtosis''), because this is precisely zero for a non-truncated lognormal distribution and positive (negative) for a distribution with fatter (thinner) tails.

In Fig.~\ref{fig.brain}, we investigate whether observed goodness-of-fit and fatness of tails are consistent with i.i.d. lognormal distributions in degree and strength, choosing the lower cutoff $x_{\min}$ for each property as the lowest half-integer which excludes the most substantial deviations from lognormality. Visual inspection reveals good fits over a substantial range but, for more quantitative insights, we consider upper-directed hypothesis tests based on the KS distance from the maximum likelihood fit to an i.i.d. lognormal process. For the degree (strength) dataset, as sample length increases, rate of rejection increases towards unity (remains around or below the test's nominal size). Therefore, according to goodness-of-fit as captured by KS distance, the degree (strength) distribution is consistent (inconsistent) with an i.i.d. lognormal distribution. However, turning to two-sided hypothesis tests employing the log-kurtosis, for both the degree and strength datasets we see rates of rejection increasing to well above nominal size as sample length increases. Therefore, in terms of the physically relevant log-kurtosis property, neither degree nor strength is consistent with an i.i.d. lognormal process. This example shows the importance of the choice of the statistic used in a statistical test and therefore  highlights the key advantage of constrained surrogates, which can be used to test an entire family of distributions based on arbitrary test statistics. 

\subsection{City populations}

Finally, we consider the paradigmatic problem of describing the distribution of the population~$x$ of the $N$ largest cities within a country. The functional form of such city-size distributions has been discussed for more than a century, with the powerlaw and lognormal models as leading candidates~\cite{malevergne2011testing,altmann2025statistical}. We apply our surrogate approaches to datasets from five countries: Australia, Brazil, Germany, the UK, and the USA. For the two largest datasets (Brazil and the USA), all five models are rejected at the 0.1 level of significance; for the three other countries, only the lognormal model avoids rejection in all cases (see Supplementary Table~\ref{tab.cities} and Fig.~\ref{fig.cities-fits}). 
These results corroborate earlier work that favored the lognormal over the powerlaw for the full distribution of US city populations~\cite{malevergne2011testing}. 

\section{Discussion and Conclusion}

%\jmm{One paragraph repeating the motivation and main story of the paper, and summarizing the goal of the paper and its main findings:}
Surrogate methods connect empirical observations with theoretical models and thus permit quantitative investigation of complex systems where analytic approaches are intractable. In this work, we introduced a general methodology for generating constrained surrogates representing entire families of continuous distributions, including some central to statistical physics and complex systems: Gaussian, lognormal, powerlaw, and exponential distributions. The resulting algorithms enable surrogate generation for arbitrary datasets, without specification or estimation of model parameters. Code to generate surrogates for arbitrary time series is available in our repository~\cite{github}. 

%\jmm{One paragraph in which we can speculate more or draw connections to other problems:}
The significance of our constrained surrogates is that they lead to improved statistical methods to:
\begin{itemize}
    \item[(i)] test the validity of an entire family of distributions $\{p_\xi(x)\}$ using arbitrary test statistics; and
\item[(ii)] obtain unbiased estimates of arbitrary statistics under the hypothesized distribution family.
\end{itemize}
These improvements derive from the fact that constrained surrogates sidestep the problem of parameter estimation to simultaneously represent a whole family of processes, thereby providing accurate hypothesis tests and unbiased estimates for arbitrary, physically meaningful quantities. 
We showed the relevance of these theoretical advantages in tests with synthetic data---Sec.~\ref{sec.performance}---and four paradigmatic applications of statistical laws in complex systems---Sec.~\ref{sec.applications}. For instance, we found that only through the use of test statistics associated with extreme events are we able to reject the hypothesis of powerlaw fire sizes in Australia, and thereby allow a more accurate assessment of the probability of extreme fires.

Which type of surrogate is preferable in practice depends on the choice of (test) statistic and, for hypothesis tests, whether the test is lower- or upper-directed.
The main practical advantage of constrained surrogates, which follows from its theoretical properties, is that they guarantee exact test size without statistic-specific benchmarking. A limitation is that constrained surrogates are necessarily uninformative for statistics fixed by the constrained properties themselves, such as maximum-likelihood estimates of model parameters. In our numerical analysis of a wide range of descriptive statistics, constrained surrogates tended to provide higher power for lower-directed tests, whereas typical surrogates higher power for upper-directed tests (see SM~\cite{supplemental}, Figs.~\ref{fig.hyp-various-stats-left-tail} and \ref{fig.hyp-various-stats-right-tail}). 
The computational cost to generate constrained surrogates is also higher and could restrict their use at large data size $N$. On a desktop computer, we estimate that a surrogate for $N=1,300$ can be generated within one second and for $N=67,000$ within  one minute. The computational time is roughly proportional to the number of MCMC transitions $\tilde{r} = O(N\log N)$ [see Eq.~(\ref{eq.rtilde})] with the constant of proportionality depending mainly on the null hypothesis and, to a lesser extent, on the input data (see Fig.~\ref{fig.comp-cost-vs-N} for the computational cost needed in different settings). 
%Timing both surrogate methods on a desktop computer across sample lengths, with $2\tilde{r}$ transitions per constrained surrogate, shows that constrained surrogates take longer to generate for all but the smallest sample lengths (see Fig.~\ref{fig.comp-cost-vs-N}). 
%For constrained surrogates, 
%computational time is roughly proportional to the number of transitions, with the constant of proportionality depending mainly on the null hypothesis and, to a lesser extent, on the input data.
%Extrapolating using the largest fitted constant of proportionality suggests that a surrogate of length

%\jmm{One paragraph about open problems, challenges, future work:}
More broadly, our constrained surrogates provide a parameter-agnostic framework for the analysis of statistical laws in complex systems that moves beyond traditional goodness-of-fit testing. By enabling hypothesis tests and estimation based on arbitrary statistics, this approach offers a principled way to distinguish agreement in some aspects of a distribution from systematic discrepancies in others. As such, constrained surrogates contribute directly to ongoing debates about the interpretation and applicability of statistical laws in empirical systems~\cite{stumpf2012critical,holme2019rare,altmann2025statistical}. Important directions for future work include applying the present framework to additional distribution families, extending it from point to interval estimation, and relaxing the i.i.d. assumption to incorporate correlations and temporal dependencies~\cite{gerlach2019testing,moore2022nonparametric}.

\section*{Acknowledgments}

\change{We gratefully acknowledge the support and hospitality of the Sydney Mathematical Research Institute (SMRI). JMM is supported by the National Natural Science Foundation of China (Grants No. No. T2541015, W2632005, and No. 12375036).}

\appendix

\section{Specific cases}\label{app.specific}

\bigskip 

We now present methods for randomizing $x_{i_1}, \ldots, x_{i_{K+1}}$ to $x'_{i_1}, \ldots, x'_{i_{K+1}}$ leading to uniform sampling among all sequences that have equal likelihood under the considered family of distributions.

\subsection{Sum}\label{ssec.sum}
Each iteration we randomize $x_{i_1}, x_{i_2}$ while maintaining the value of $\beta_1 = x_{i_1} + x_{i_2}$. We consider the change of coordinates
\begin{align}
    \beta_1 &= x_{i_1} + x_{i_2},\notag\\
    s &= x_{i_1},\notag
\end{align}
which corresponds to Jacobian
\[\left\lvert \det \pp{\bbeta, s}{{\bf x}_{K+1}} \right\rvert = \left\lvert \det \left(
\begin{array}{cc}
1 & 1\\
1 & 0
\end{array}
\right) \right\rvert = 1.\]
It follows from Eq.~(\ref{eq.prob_dens_t}) that we can achieve symmetric transition probabilities by choosing $s$ with constant probability density; i.e., uniformly on $[0, \beta_1]$.

With $s$ chosen uniformly at random, transition probabilities are symmetric. Because a transition which does not change the current state is always possible, the Markov chain defined by the algorithm is aperiodic. Therefore, the algorithm will sample uniformly at random from some set satisfying the constraint. To show that our algorithm has full range, i.e., that it samples uniformly at random from the \emph{entire} collection of sequences with the same likelihood, we show that it is possible, using the algorithm, to reach a chosen sequence ${\bf x}''$ starting from any sequence ${\bf x}$ such that $\sum x_i = \sum x''_i$. Let $x''_{i}$ be the maximum value of ${\bf x}''$, and let $x_{j}$ be the largest value of ${\bf x}$ except, possibly, for $x_{i}$. If $x_{i} + x_{j} \geq x''_{i}$ then perform the transition $x_{i}, x_{j} \rightarrow x''_{i}, x_{i} + x_{j} - x''_{i}$, otherwise, change ${\bf x}$ according to $x_{i}, x_{j} \rightarrow x_{i} + x_{j}, 0$. After several repeats of this operation, we will have $x_i = x''_i$. The sequences ${\bf x}$ and ${\bf x}''$ now fail to match in at most $N-1$ places and, repeating the argument, we are able to iteratively reduce the number of mismatches until ${\bf x}$ and ${\bf x}''$ coincide at every point.

\subsection{Sum and second moment}

Each iteration we randomize $x_{i_1}, x_{i_2}, x_{i_3}$ while maintaining the values of $\beta_1 = x_{i_1} + x_{i_2} + x_{i_3}$ and $\beta_2 = {x_{i_1}}^2 + {x_{i_2}}^2 + {x_{i_3}}^2$. The second condition describes a sphere in $\mathbb{R}^3$ with center at the origin and radius $\sqrt{\beta_2}$, while the first represents a plane with normal $\hat{\bf n} = \frac{1}{\sqrt{3}}(1, 1, 1)^T$ and passing through the point ${\bf c} = \frac{\beta_1}{3}(1, 1, 1)^T$. The set of allowable points lie on the intersection of these surfaces, which is a circle with center ${\bf c}$ and radius which can be identified as
\begin{align}
    \rho \left( \bbeta \right) = \frac{1}{\sqrt{3}} \sqrt{3 \beta_2 - {\beta_1}^2}.\notag
\end{align}

We can transform from a coordinate system $(x, y, z)$ in which this circle is centered at the origin and lies in the $xy$-plane via
\begin{align}
    \vecto{x_{i_1}}{x_{i_2}}{x_{i_3}} = R \cdot \vecto{x}{y}{0} + \frac{1}{\sqrt{3}}\vecto{z}{z}{z},\notag
\end{align}
where
\begin{align}
    R = \left(
    \begin{matrix}
    0 & \sqrt{\frac{2}{3}} & \frac{1}{\sqrt{3}}\\
    -\frac{1}{\sqrt{2}} & -\frac{1}{\sqrt{6}} & \frac{1}{\sqrt{3}}\\
    \frac{1}{\sqrt{2}} & -\frac{1}{\sqrt{6}} & \frac{1}{\sqrt{3}}
    \end{matrix}
    \right)\notag
\end{align}
is a rotation matrix. The rotation matrix $R$ has been identified by choosing its first two columns, to which $R$ maps unit vectors in the $x$- and $y$-directions, as orthonormal vectors in the plane normal to $\hat{\bf n}$, and choosing its third column as $\hat{\bf n}$ itself. The first two columns are ordered such that the cross product of the first with the second is $\hat{\bf n}$ (rather than $-\hat{\bf n}$), which leads to a determinant of $+1$ (rather than $-1$). Moving to coordinates $\bbeta, s$ such that
\begin{align}
    \vecto{u}{v}{z} = \vecto{\rho(\bbeta) \cos s}{\rho(\bbeta) \sin s}{\beta_1/\sqrt{3}},\notag
\end{align}
we reduce randomization of $\left( x_{i_1}, x_{i_2}, x_{i_3}\right)^T$ to a choice of $s \in [0, 2\pi)$.

To determine the appropriate distribution for $s$, we need to know how the Jacobian $\left\lvert \det \pp{\bbeta, s}{{\bf x}_{K+1}} \right\rvert$ depends on $s$. To determine this, we note that the Jacobian of the inverse transformation can be written
\begin{align}
    \left\lvert \det \pp{{\bf x}_{K+1}}{\bbeta, s} \right\rvert = \left\lvert \det \pp{{\bf x}_{K+1}}{x, y, z} \right\rvert \left\lvert \det \pp{x, y, z}{\bbeta, s} \right\rvert. \notag
\end{align}
The first factor is
\begin{align}
\left\lvert \det \pp{{\bf x}_{K+1}}{x, y, z} \right\rvert =& \left\lvert \det R \right\rvert = 1,\notag
\end{align}
while the second factor is
\begin{align}
    \left\lvert \det \pp{x, y, z}{\bbeta, s} \right\rvert =& \left\lvert \det \left( \begin{matrix}
        \pp{\rho}{\beta_1} \cos s & \pp{\rho}{\beta_2} \cos s & -\rho(\bbeta)\sin s\\
        \pp{\rho}{\beta_1} \sin s & \pp{\rho}{\beta_2} \sin s & \rho(\bbeta)\cos s\\
        \frac{1}{\sqrt{3}} & 0 & 0
    \end{matrix}\right)  \right\rvert \notag\\
    =&\frac{1}{\sqrt{3}} \rho(\bbeta) \pp{\rho}{\beta_2}.\notag
\end{align}
Their product is therefore $\left\lvert \det \pp{{\bf x}_{K+1}}{\bbeta, s} \right\rvert = \frac{1}{\sqrt{3}} \rho(\bbeta) \pp{\rho}{\beta_2}$. Since this is independent of $s$, so is its inverse, the Jacobian $\left\lvert \det \pp{\bbeta, s}{{\bf x}_{K+1}} \right\rvert$. It follows from Eq.~(\ref{eq.prob_dens_t}) that we can achieve symmetric transition probabilities by choosing $s$ uniformly on $[0, 2\pi)$.

Arguments similar to those for constrained exponential surrogates (See Sec.~\ref{ssec.sum}) imply aperiodicity and full range.

\subsection{Product} \label{ssec.product}

To generate constrained powerlaw surrogates we must randomize a sequence ${\bf x}$ while fixing the product $\prod_i {x_i}$. We consider two theoretically equivalent approaches, the first of which is more direct. Each iteration we randomize $x_{i_1}, x_{i_2}$ while maintaining the value of $\beta_1 = x_{i_1} x_{i_2}$. We consider the change of coordinates
\begin{align}
    \beta_1 &= x_{i_1} x_{i_2},\notag\\
    s &= x_{i_1},\notag
\end{align}
which corresponds to Jacobian
\[\left\lvert \det \pp{\bbeta, s}{{\bf x}_{K+1}} \right\rvert = \left\lvert \det \left(
\begin{array}{cc}
x_{i_2} & x_{i_1}\\
1 & 0
\end{array}
\right) \right\rvert = x_{i_1} = s.\]
It follows from Eq.~(\ref{eq.prob_dens_t}) that we can achieve symmetric transition probabilities by choosing $s \in [1, \beta_1]$ with probability density $P(s) = C/s$. A change of variables shows that this is equivalent to choosing $\log s$ uniformly on $[0, \log \beta_1]$, which suggests our second approach to generating constrained powerlaw surrogates.

This alternative method for generating constrained powerlaw surrogates allows us to utilize the existing solution for generating constrained exponential surrogates. Since fixing the product $\prod_i {x_i}$ is equivalent to fixing $\sum_i \log x_i$, we apply our constrained exponential surrogate algorithm to the log-transformed sequence ${\bf y} = \left( y_1, \ldots, y_N \right)$ where, for $i = 1, \ldots, N$, $y_i = \log x_i$, then exponentiate the resulting sequence. Our constrained powerlaw surrogate algorithm inherits the full range of the constrained exponential surrogate algorithm. To show uniform sampling we need to show that the probability $P_{\bf X}({\bf x})$ of observing a sequence ${\bf x}$ after applying our constrained powerlaw surrogate algorithm satisfies
\begin{align}
P_{\bf X}({\bf x}) = P_{\bf X}({\bf x}')\label{eq.power_eq_prob}
\end{align}
for any pair of sequences ${\bf x}, {\bf x}'$ which have the same product. We already know that the analogous result \begin{align}
    P_{\bf Y}({\bf y}) = P_{\bf Y}({\bf y}') \label{eq.expon_eq_prob}
\end{align}
holds for ${\bf y}, {\bf y}'$. For the change of variables ${\bf y} = {\bf f} \left( {\bf x} \right)$ we have Jacobian $\left\lvert {\bf f}' \left( {\bf x} \right) \right\rvert = 1/\prod_i x_i$, and so Eq.~(\ref{eq.expon_eq_prob}) transforms to $P_{\bf X}({\bf x}) \prod_i x_i = P_{\bf X}({\bf x}') \prod_i x'_i$. Equation~(\ref{eq.power_eq_prob}), which implies uniform sampling, now follows from preservation of the product $\prod_i x_i$.

From this argument we infer a slightly more general strategy. To produce a constrained surrogate for $x_1, \ldots, x_N$ while fixing $\beta_1 \left( x_1, \ldots, x_N \right) = \prod\limits_{i=1}^N x_i, \beta_2 \left( x_1, \ldots, x_N \right), \ldots, \beta_K \left( x_1, \ldots, x_N \right)$, we can take a surrogate of its log-transform ${\bf y}$ which preserves $\beta_1 \left( \exp y_1, \ldots, \exp y_N \right), \ldots, \beta_K \left( \exp y_1, \ldots, \exp y_N \right)$, then perform an inverse log transform on the result. The next example is of this this type.

\subsection{Product and second log-moment}\label{ssec.product and second moment}

To generate constrained lognormal surrogates we must randomize a sequence ${\bf x}$ while fixing the first and second log-moments $\sum_i \log x_i$ and $\sum_i \left(\log x_i \right)^2$. To accommodate these constraints, we apply our constrained truncated Gaussian surrogate algorithm to the log-transformed sequence ${\bf y}$, then exponentiate the resulting sequence. Our constrained lognormal surrogate algorithm inherits from our constrained truncated Gaussian surrogate algorithm full range and, by preservation of the product $\prod_i x_i$, uniform sampling (For analogous reasoning, see Sec.~\ref{ssec.product}).

\subsection{Maximum}\label{ssec.maximum}
To randomize the sequence ${\bf x}$ while retaining the maximum $\beta_1 = \max\limits_{i \in \{1,\ldots,N\}} x_{i}$, we independently choose each element $x_i$ uniformly at random from the interval $[x_{\min}, \beta_1)$. We then uniformly at randomly choose a single element $x_i$ and set its value to preserve the originally observed maximum of the sequence, i.e., as $x_i = \beta_1$. Note that we have constrained the property which determines the likelihood of a uniform distribution without applying the majority of our framework. 

\section{Number of transitions}\label{app.transitions}

Here we estimate the number of transitions required to approximate the final distribution. Our main finding is that for any $p \in [1,\infty]$ and tolerance $\epsilon > 0$, the relative error in expectation after $r$ iterations will satisfy
\begin{equation}
 \frac{\left\lVert \mathbb{E}\left[\xvecit{r}\right] - \mathbb{E}\left[\xvecit{\infty}\right]\right\rVert_p}{\left\lVert\mathbb{E}\left[\xvecit{\infty}\right]\right\rVert_p} < \epsilon \label{eq.error}
\end{equation}
as long as $r \geq \frac{N}{b} \left(\log_b N - \log_b (\epsilon)\right)$, where $b=2$ ($b=3$) for constrained exponential (truncated Gaussian) surrogates and, for constrained powerlaw (lognormal) surrogates, the same result holds for the continuous transformation $\log {x^{(0)}}_1, \ldots, \log {x^{(0)}}_N$ of the observed time series $\xvecit{0} = {x^{(0)}}_1, \ldots, {x^{(0)}}_N$. In this expression, $\xvecit{r} = \xit{r}_1, \ldots, \xit{r}_N$ represents the result of $r$ iterations of a constrained surrogate algorithm, and $\mathbb{E}[\cdot]$ is the expectation over many independent trials. In particular, $\mathbb{E}\left[\xvecit{0}\right] = \xvecit{0}$---the originally observed time series---and $\mathbb{E}\left[\xvecit{\infty}\right] := 
\lim\limits_{r\rightarrow\infty} \mathbb{E}\left[\xvecit{r}\right] = m, \ldots, m$ comprises $N$ instances of the mean value $m$ of the limiting expectation $\mathbb{E}\left[\xvecit{\infty}\right]$.

To demonstrate this, we first consider the exponential surrogate algorithm and the case $x_{\min} = 0$. We investigate the observed sequence $\xvecit{0} = Nm, 0, 0, \ldots, 0$, which represents the most extreme deviation from the limiting expectation $\mathbb{E}\left[\xvecit{\infty}\right]$, and assume that any other observed sequence will achieve condition (\ref{eq.error}) at least as quickly. We also assume that for $r \geq 0$ and $t \geq 2$,
\begin{align}
    \mathbb{E} \left[ \xit{r}_1 \right] \geq m \geq \mathbb{E} \left[ \xit{r}_t \right].\label{eq.xt}
\end{align}
For an iteration of the constrained exponential surrogate algorithm which involves $\xit{r}_1$ and $\xit{r}_t$, we have
\begin{align}
    \mathbb{E} \left[ \xit{r + 1}_1 \right] =& \frac{1}{2} \left( \mathbb{E} \left[ \xit{r}_1\right] + \mathbb{E} \left[ \xit{r}_t\right] \right) \leq \frac{1}{2} \mathbb{E} \left[ \xit{r}_1 \right] + \frac{m}{2},\notag    
\end{align}
where the inequality follows from the right hand inequality in relationship (\ref{eq.xt}). Hence the error $\mathbb{E} \left[ \xit{r}_1 \right] - \mathbb{E} \left[ \xit{\infty}_1 \right] = \mathbb{E} \left[ \xit{r}_1 \right] - m$ in the expectation of $\xit{r}_1$ satisfies
\begin{align}
    \left\lvert \mathbb{E} \left[ \xit{r + 1}_1 \right]  - m \right\rvert \leq \frac{1}{2} \left\lvert \mathbb{E} \left[ \xit{r}_1 \right] - m \right\rvert,\notag
\end{align}
where the absolute value can be used because of the left hand inequality in relationship (\ref{eq.xt}). This implies that the error in $\mathbb{E} \left[ \xit{r}_1 \right]$ more than halves in each iteration which involves $\xit{r}_t$. Since, on average, $\xit{r}_t$ will be involved in $2$ in $N$ iterations, we can expect error to shrink as
\begin{align}
    \left\lvert \mathbb{E} \left[ \xit{r}_1 \right]  - m \right\rvert \leq& 2^{-2r/N} \left\lvert \mathbb{E} \left[ \xit{0}_1 \right]  - m \right\rvert\notag\\
    &= 2^{-2r/N} \left\lvert N m  - m \right\rvert \leq 2^{-2r/N} N m.\notag
\end{align}
The same argument holds for other coordinates, so we have that for $t \geq 1$,
\begin{align}
    \left\lvert \mathbb{E} \left[ \xit{r}_t \right]  - m \right\rvert \leq 2^{-2r/N} N m.\notag
\end{align}
The relative error in $\mathbb{E} \left[ \xvecit{r} \right]$ must then satisfy
\begin{align}
    \frac{\left\lVert \mathbb{E} \left[ \xvecit{r} \right]  - \mathbb{E} \left[ \xvecit{\infty} \right] \right\rVert_p}{\left\lVert \mathbb{E} \left[ \xvecit{\infty} \right] \right\rVert_p} \leq \frac{N^{1/p} 2^{-2r/N} N m}{N^{1/p} m} = 2^{-2r/N} N.\notag
\end{align}
This holds also for $x_{\min} > 0$, as does the tighter bound
\begin{align}
    \frac{\left\lVert \mathbb{E} \left[ \xvecit{r} \right]  - \mathbb{E} \left[ \xvecit{\infty} \right] \right\rVert_p}{\left\lVert \mathbb{E} \left[ \xvecit{\infty} \right] \right\rVert_p} \leq 2^{-2r/N} N \frac{m - x_{\min}}{m}.\notag
\end{align}
Therefore, for constrained exponential surrogates, relative error less than $\epsilon$ is expected for a number of iterations $r \geq \frac{N}{2} \left( \log_2 N - \log_2 \epsilon \right)$.

For an iteration of the constrained truncated Gaussian surrogate algorithm which involves $\xit{r}_1$, $\xit{r}_t$ and $\xit{r}_u$, we instead have
\begin{align}
    \mathbb{E} \left[ \xit{r + 1}_1 \right] =& \frac{1}{3} \left( \mathbb{E} \left[ \xit{r}_1\right] + \mathbb{E} \left[ \xit{r}_t\right] + \mathbb{E} \left[ \xit{r}_u\right] \right)\notag    
\end{align}
and expect $\xit{r}_t$ to be involved in $3$ in $N$ iterations. The same arguments as for constrained exponential surrogates indicate that, for the constrained truncated Gaussian surrogate algorithm, we can achieve relative error less than $\epsilon$ for a number of iterations $r \geq \frac{N}{3} \left( \log_3 N - \log_3 \epsilon \right)$.

\bibliography{ref}% Produces the bibliography via BibTeX.

\onecolumngrid

\clearpage

% \section*{Supplemental material}

\begin{center}
{\large\bfseries Supplemental material}

\vspace{0.8em}
{\large for}

\vspace{0.8em}
{\large\bfseries Constrained surrogates for arbitrary families of continuous probability distributions}

\vspace{1.2em}
Jack Murdoch Moore and Eduardo G. Altmann
\end{center}

\vspace{0.8em}

\setcounter{page}{1}

\setcounter{section}{0}
\setcounter{figure}{0}
\setcounter{table}{0}
\renewcommand\thefigure{S\arabic{figure}}
\renewcommand\thetable{S\arabic{table}}

\begin{figure*}[h]
    \includegraphics[width=\textwidth]{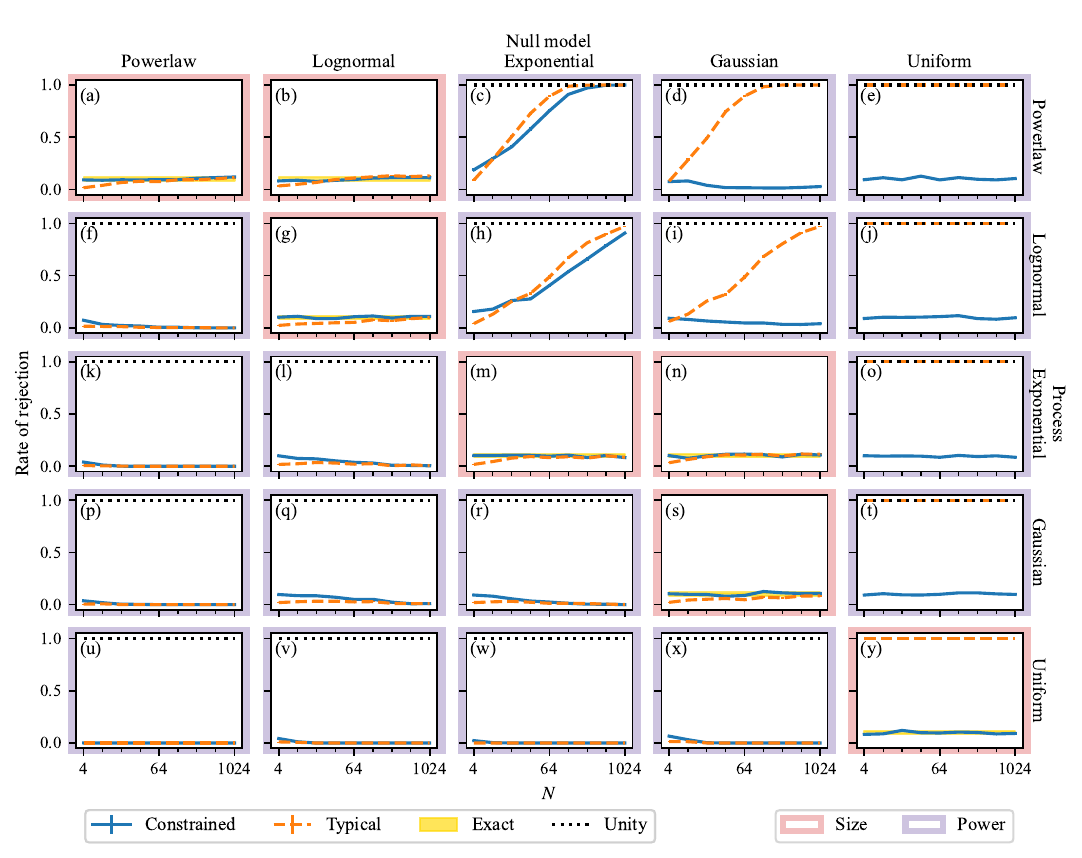}
    \caption{Analogue of Fig.~\ref{fig.hyp-max-left-tail} for upper-directed hypothesis tests, also using the maximum as test statistic. Rates of rejection estimated from $1,000$ hypothesis tests, each using an independent sample of length $N$ i.i.d. from a (from top to bottom): powerlaw; lognormal; exponential; truncated Gaussian; and uniform distribution. Each test uses 19 constrained (blue solid) or typical (orange solid) surrogates representing the null hypothesis of a (from left to right): (a-e) powerlaw; (f-j) lognormal; (k-o) exponential; (p-t) truncated Gaussian; and (u-y) uniform distribution. Tests are lower-directed, have nominal size 10\%, and use as the sample maximum as discriminating statistic. Panels showing tests for which the null hypothesis is correct (i.e., for which the rate of rejection is the size of the test) have red borders, and feature a gold band spanning the 90\% confidence interval for a Bernoulli process. Panels showing tests for which the null hypothesis is incorrect  (i.e., for which the rate of rejection is the test's power to reject an incorrect hypothesis) are given purple borders, and feature a black line corresponding to the ideal (100\%) rejection rate. Error bars correspond to standard error but only slightly exceed the line width.
    }
    \label{fig.hyp-max-right-tail}
\end{figure*}

\begin{figure*}
    \includegraphics[width=\textwidth]{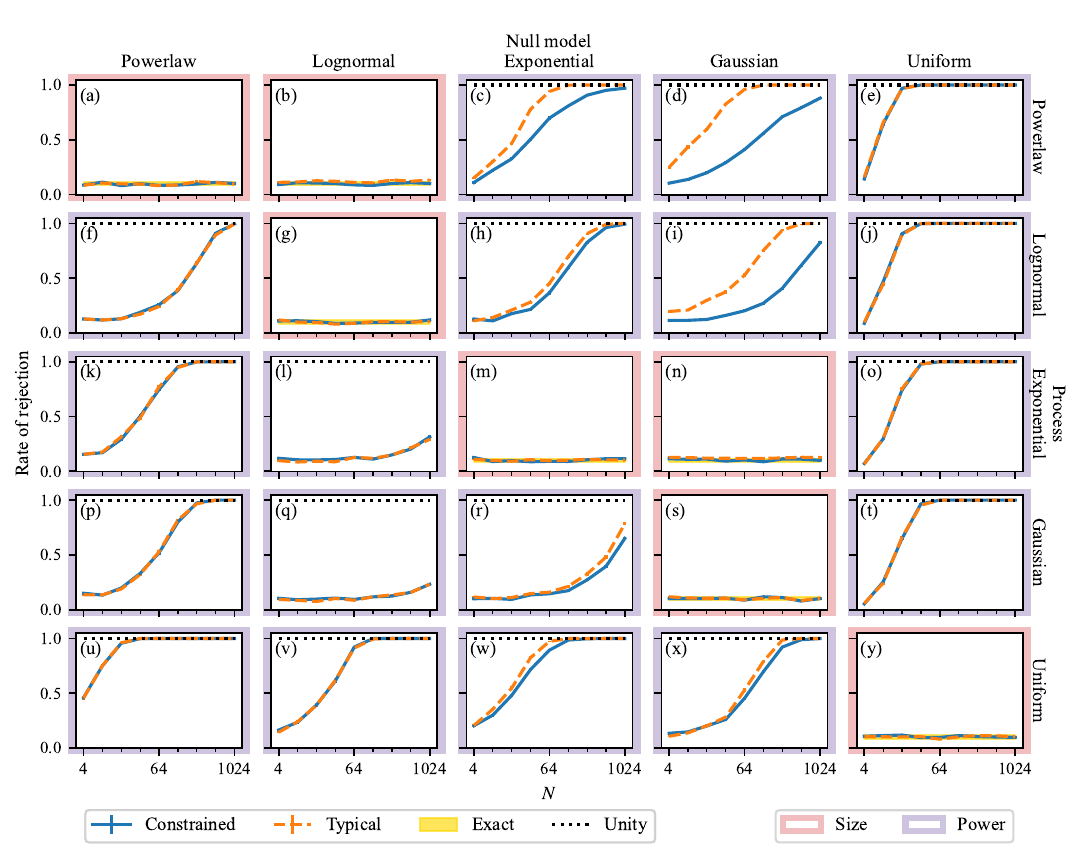}
    \caption{Analogue of Fig.~\ref{fig.hyp-max-left-tail} for upper-directed hypothesis tests using the KS distance as test statistic. Rates of rejection estimated from $1,000$ hypothesis tests, each using an independent sample of length $N$ i.i.d. from a (from top to bottom): powerlaw; lognormal; exponential; truncated Gaussian; and uniform distribution. Each test uses 19 constrained (blue solid) or typical (orange solid) surrogates representing the null hypothesis of a (from left to right): (a-e) powerlaw; (f-j) lognormal; (k-o) exponential; (p-t) truncated Gaussian; and (u-y) uniform distribution. Tests are lower-directed, have nominal size 10\%, and use as the sample maximum as discriminating statistic. Panels showing tests for which the null hypothesis is correct (i.e., for which the rate of rejection is the size of the test) have red borders, and feature a gold band spanning the 90\% confidence interval for a Bernoulli process. Panels showing tests for which the null hypothesis is incorrect  (i.e., for which the rate of rejection is the test's power to reject an incorrect hypothesis) are given purple borders, and feature a black line corresponding to the ideal (100\%) rejection rate. Error bars correspond to standard error but only slightly exceed the line width.
    }
    \label{fig.hyp-KS-dist}
\end{figure*}

\begin{figure*}
    \includegraphics[width=\textwidth]{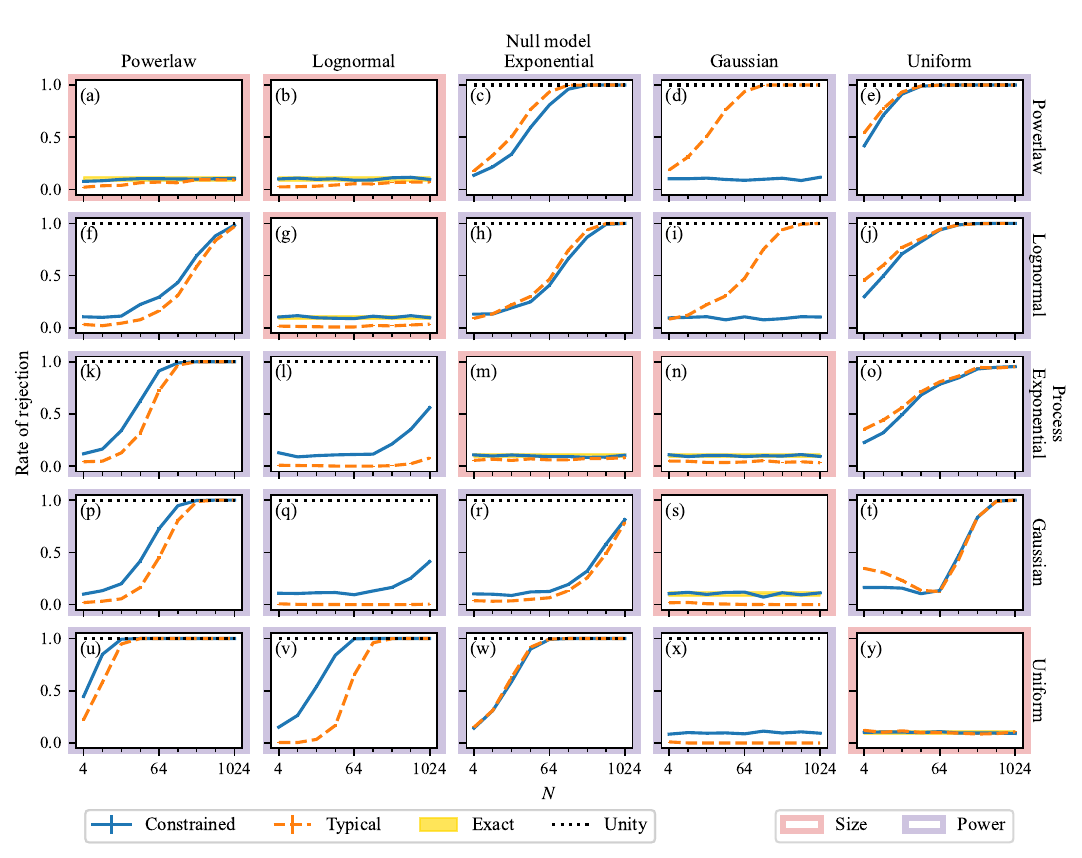}
    \caption{Analogue of Fig.~\ref{fig.hyp-max-left-tail} for two-sided hypothesis tests using the coefficient of variation $\sigma_x/\mu_x$ as test statistic. Rates of rejection estimated from $1,000$ hypothesis tests, each using an independent sample of length $N$ i.i.d. from a (from top to bottom): powerlaw; lognormal; exponential; truncated Gaussian; and uniform distribution. Each test uses 19 constrained (blue solid) or typical (orange solid) surrogates representing the null hypothesis of a (from left to right): (a-e) powerlaw; (f-j) lognormal; (k-o) exponential; (p-t) truncated Gaussian; and (u-y) uniform distribution. Tests are lower-directed, have nominal size 10\%, and use as the sample maximum as discriminating statistic. Panels showing tests for which the null hypothesis is correct (i.e., for which the rate of rejection is the size of the test) have red borders, and feature a gold band spanning the 90\% confidence interval for a Bernoulli process. Panels showing tests for which the null hypothesis is incorrect  (i.e., for which the rate of rejection is the test's power to reject an incorrect hypothesis) are given purple borders, and feature a black line corresponding to the ideal (100\%) rejection rate. Error bars correspond to standard error but only slightly exceed the line width.
    }
    \label{fig.hyp-coef-var}
\end{figure*}

\begin{figure*}
    \includegraphics[width=\textwidth]{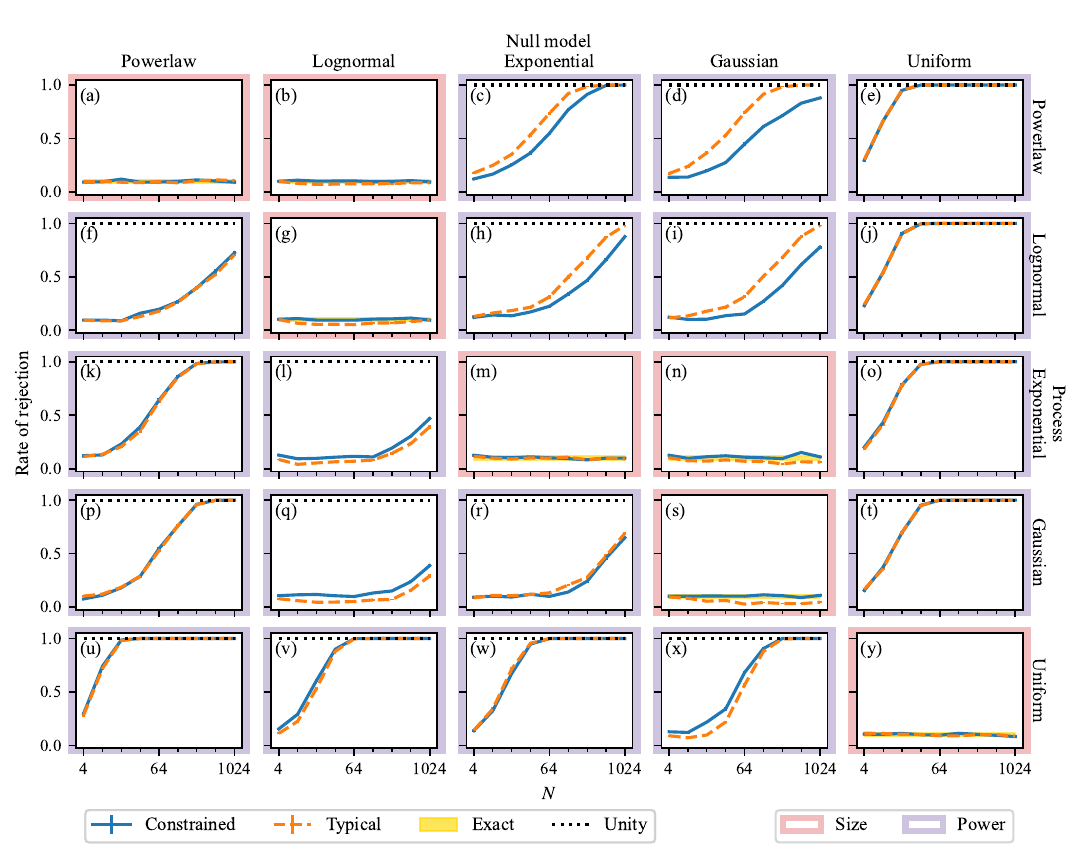}
    \caption{Analogue of Fig.~\ref{fig.hyp-max-left-tail} for two-sided hypothesis tests using log-skewness as test statistic. Rates of rejection estimated from $1,000$ hypothesis tests, each using an independent sample of length $N$ each drawn i.i.d. from a: (a-e) powerlaw; (f-j) lognormal; (k-o) exponential; (p-t) truncated Gaussian; and (u-y) uniform distribution. Each test uses 19 constrained (blue solid) or typical (orange solid) surrogates representing the true underlying process and correct null model of a (from left to right): powerlaw, lognormal, exponential, truncated Gaussian, and uniform distribution. Tests have nominal size 10\%, and use as discriminating statistic the KS-distance. The gold band spans the 90\% confidence interval for a Bernoulli process, and the error bars of other lines correspond to standard error but only slightly exceed the line width.
    }
    \label{fig.hyp-skewness}
\end{figure*}

\begin{figure}
    \includegraphics[width=\textwidth]{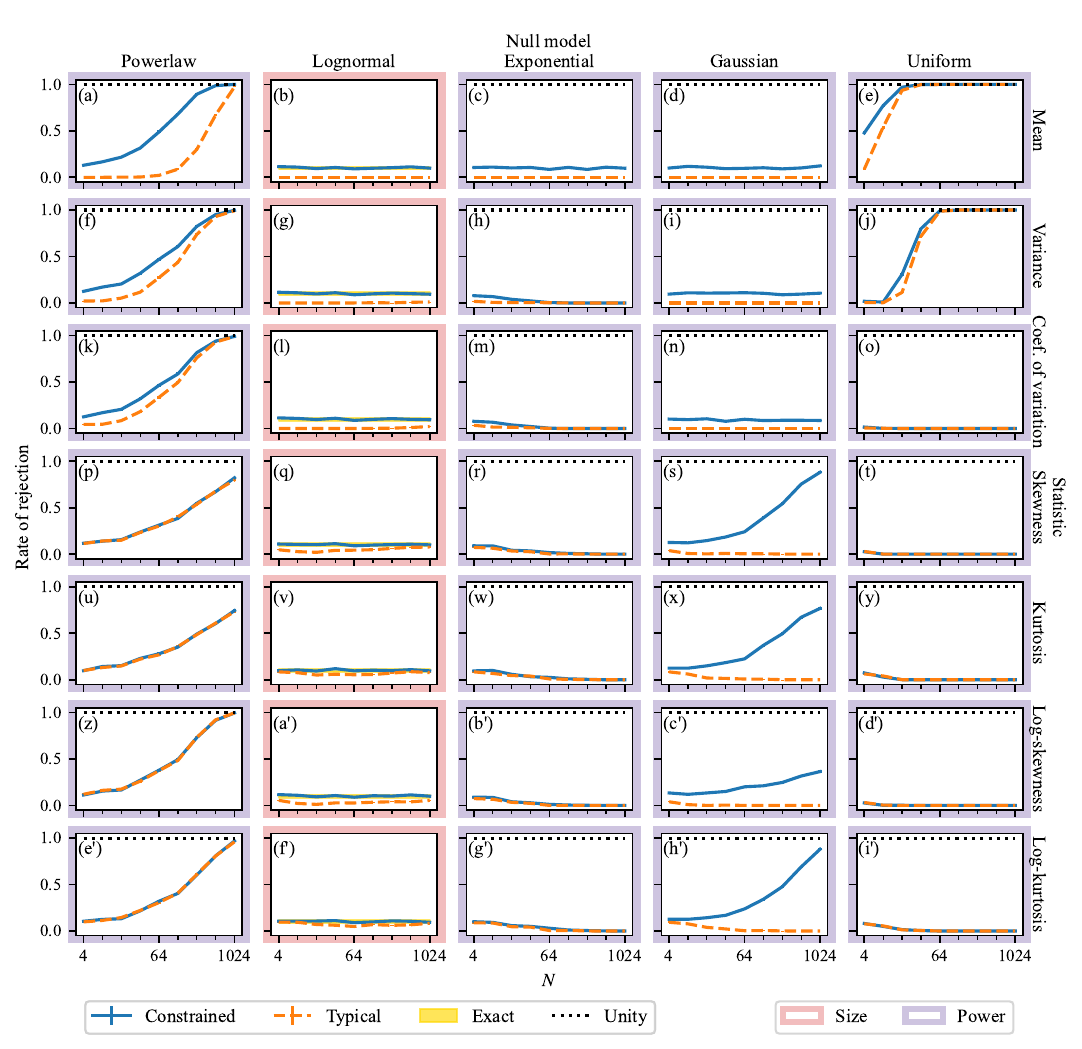}
    \caption{For lower-directed hypothesis tests, constrained surrogates outperform typical surrogates across a wide range of descriptive discriminating statistics. Rates of rejection estimated from $1,000$ hypothesis tests, each using an independent sample of length $N$ i.i.d. from a lognormal distribution. Each test uses 19 constrained (blue solid) or typical (orange solid) surrogates representing the null hypothesis of a (from left to right): powerlaw; lognormal; exponential; truncated Gaussian; and uniform distribution. Tests are lower-directed, have nominal size 10\%, and use as discriminating statistic (from top to bottom): mean; variance; coefficient of variation; skewness; kurtosis; skewness of the log transformed data (log-skewness); and kurtosis of the log transformed data (log-kurtosis). Panels showing tests for which the null hypothesis is correct (i.e., for which the rate of rejection is the size of the test) have red borders, and feature a gold band spanning the 90\% confidence interval for a Bernoulli process. Panels showing tests for which the null hypothesis is incorrect  (i.e., for which the rate of rejection is the test's power to reject an incorrect hypothesis) are given purple borders, and feature a black line corresponding to the ideal (100\%) rejection rate. Error bars correspond to standard error but only slightly exceed the line width.
    }
    \label{fig.hyp-various-stats-left-tail}
\end{figure}

\begin{figure}
    \includegraphics[width=\textwidth]{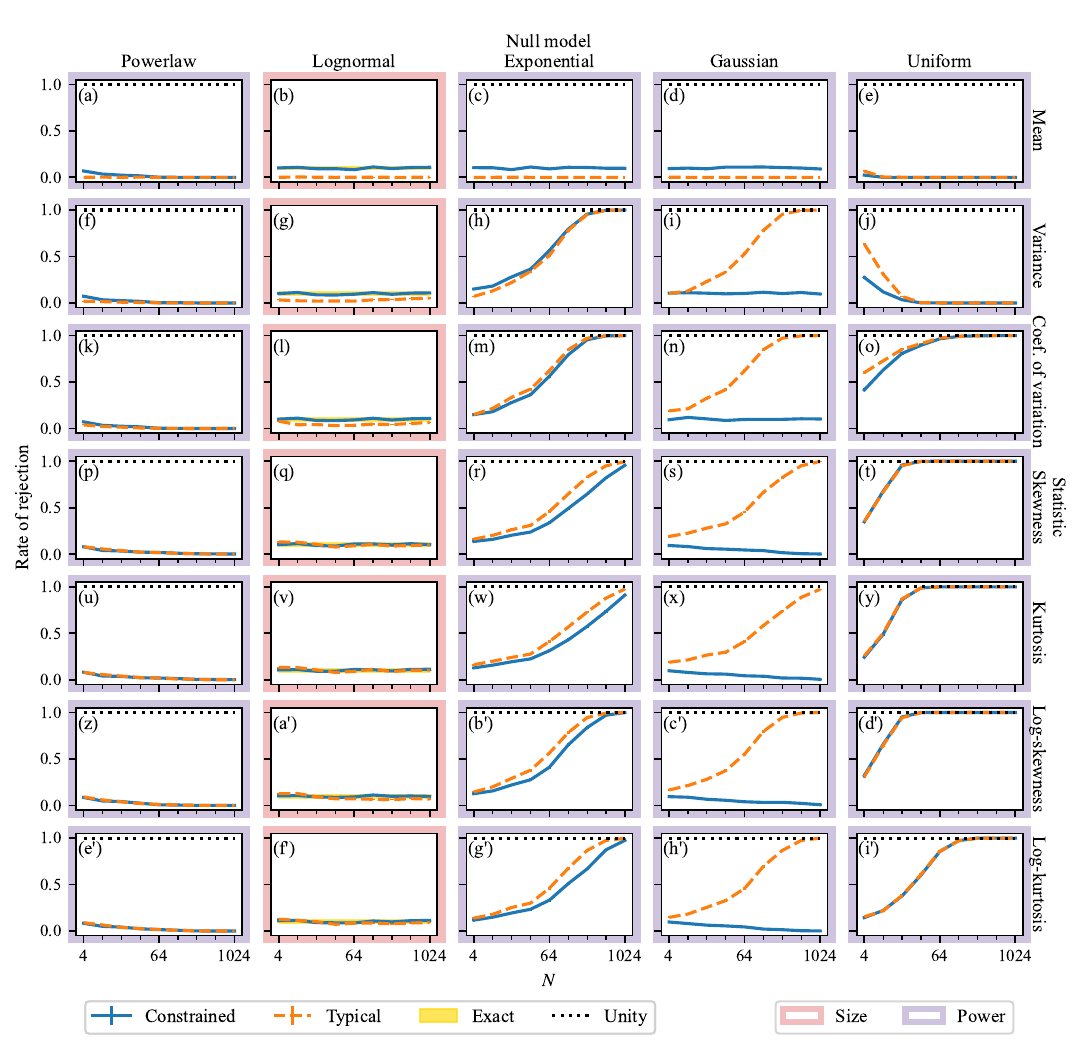}
    \caption{For upper-directed hypothesis tests across a wide range of descriptive discriminating statistics, typical surrogates can provide higher power than can constrained surrogates. Rates of rejection estimated from $1,000$ hypothesis tests, each using an independent sample of length $N$ i.i.d. from a lognormal distribution. Each test uses 19 constrained (blue solid) or typical (orange solid) surrogates representing the null hypothesis of a (from left to right): powerlaw; lognormal; exponential; truncated Gaussian; and uniform distribution. Tests are upper-directed, have nominal size 10\%, and use as discriminating statistic (from top to bottom): mean; variance; coefficient of variation; skewness; kurtosis; skewness of the log transformed data (log-skewness); and kurtosis of the log transformed data (log-kurtosis). Panels showing tests for which the null hypothesis is correct (i.e., for which the rate of rejection is the size of the test) have red borders, and feature a gold band spanning the 90\% confidence interval for a Bernoulli process. Panels showing tests for which the null hypothesis is incorrect  (i.e., for which the rate of rejection is the test's power to reject an incorrect hypothesis) are given purple borders, and feature a black line corresponding to the ideal (100\%) rejection rate. Error bars correspond to standard error but only slightly exceed the line width.
    }
    \label{fig.hyp-various-stats-right-tail}
\end{figure}

\begin{figure*}[h]
    \includegraphics[width=\textwidth]{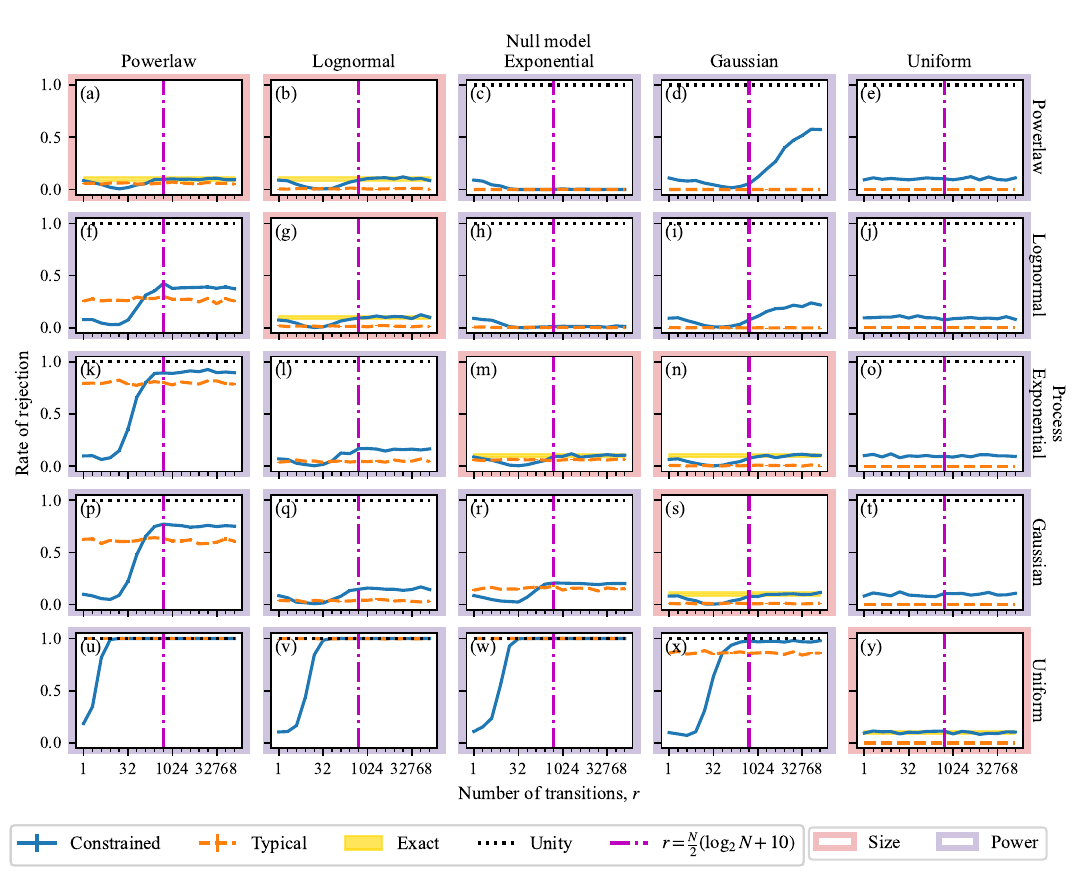}
    \caption{Rates of rejection have usually plateaued by the number of transitions per surrogate considered in the main text. Rates of rejection when each constrained surrogate is generated independently from the original observed sequence instead of from the previously generated constrained surrogate. Results show the average over $1000$ hypothesis tests, each using an independent sample of length $N = 64$ i.i.d. from a (from top to bottom): powerlaw; lognormal; exponential; truncated Gaussian; and uniform distribution. Each test uses 19 constrained (blue solid) or typical (orange solid) surrogates representing the null hypothesis of a (from left to right): (a-e) powerlaw; (f-j) lognormal; (k-o) exponential; (p-t) truncated Gaussian; and (u-y) uniform distribution. Tests are lower-directed, have nominal size 10\%, and use as the sample maximum as discriminating statistic. Panels showing tests for which the null hypothesis is correct (i.e., for which the rate of rejection is the size of the test) have red borders, and feature a gold band spanning the 90\% confidence interval for a Bernoulli process. Panels showing tests for which the null hypothesis is incorrect  (i.e., for which the rate of rejection is the test's power to reject an incorrect hypothesis) are given purple borders, and feature a black line corresponding to the ideal (100\%) rejection rate. A vertical line (magenta dot-dashed) shows where the number of transitions is $r = \tilde{r} = \frac{N}{2}\left(\log_2 N - \log_2 \epsilon \right)$, which is expected to achieve error in expectation less than $\epsilon = 10^{-10}$. Error bars correspond to standard error but only slightly exceed the line width.
    }
    \label{fig.vs-nt-hyp-max-left-tail}
\end{figure*}

\begin{figure*}[htbp]
    \centering
    \includegraphics[width=\textwidth]{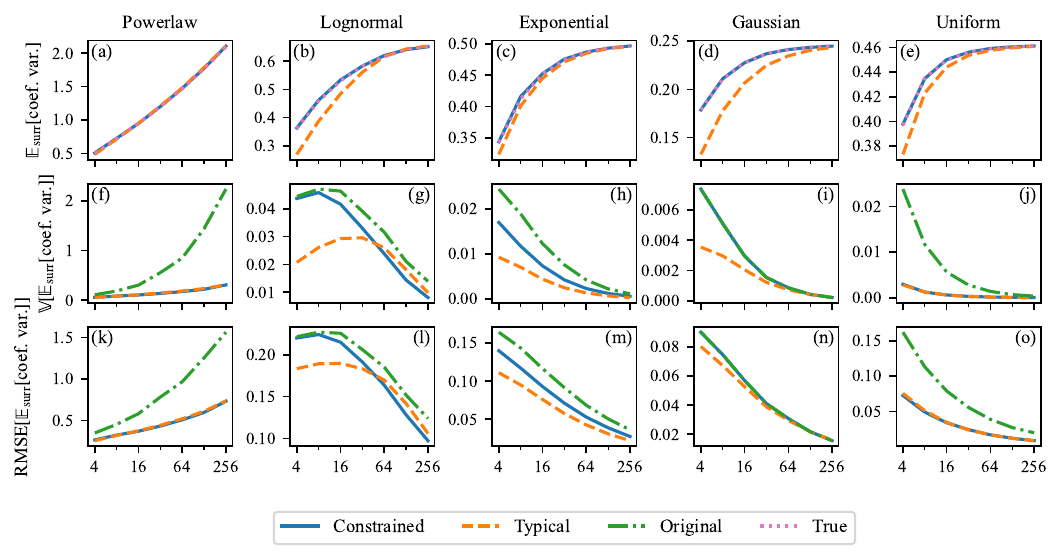}
    \caption{%
    Analogue of Fig.~\ref{fig.estimation}, but considering, rather than maximum, the coefficient of variation $\sigma_x/\mu_x$. Constrained surrogates do not bias sample statistics, and can reduce variance and error. Expectation (first row), variance $\mathbb{V}$ (second row), and root mean square error (RMSE, third row) of estimated expectation $\mathbb{E}_{\text{surr}}$ of sample coefficient of variation. $\mathbb{V}$ is calculated using $10^4$ independent input sequences ${\bf x^{\rm{input}}}$, each comprising $N$ values drawn i.i.d. from a (from left to right): powerlaw, lognormal, exponential, truncated Gaussian, and uniform distribution.
    $\mathbb{E}_{\text{surr}}$ is estimated using $100$ surrogates, each from the same ${\bf x^{\rm{input}}}$. The surrogates considered are either constrained (blue solid), typical (orange dashed), unchanged from the original observation (green dot-dashed), or generated independently from the true underlying process (pink dotted).
    }
    \label{fig.estimation-coef-var}
\end{figure*}

\begin{figure*}
\includegraphics[width=\textwidth]{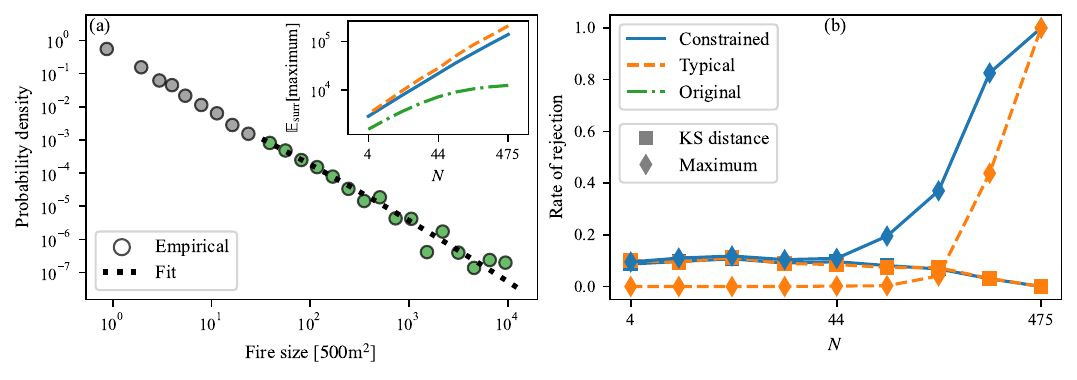}
    \caption{Analogue of Fig.~\ref{fig.fires} with additional combinations of surrogate type and test statistic. (a) Log-binned empirical distribution of fire magnitude (circles) and the maximum likelihood fit to a lognormal model (dotted black line) of values above the lower cutoff. Inset: Expectations of sample maximum estimated as the mean over $10^5$ independent samples of length $N$ downsampled (i.e., drawn i.i.d. without replacement) either from the original data or surrogates produced from the full dataset. (b) Rates of rejection estimated from 1,000 hypothesis tests, each using a sample of length $N$ from the observed dataset, 99 surrogates, and either the KS distance or maximum. Only values of the original data which exceed the fitted lower cutoff $x_{\min} = 32.5\ [500 \textrm{m}^2]$ are considered.
    }
    \label{fig.fires-supp}
\end{figure*}

\begin{figure}
\includegraphics[width=\textwidth]{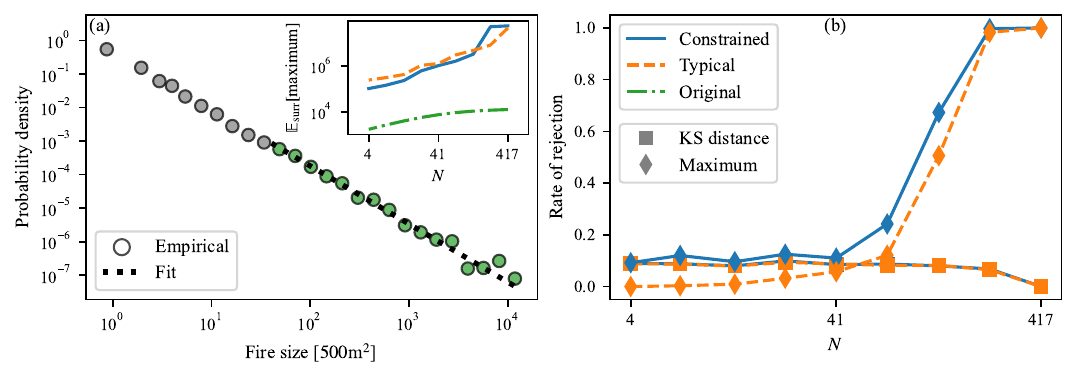}
    \caption{Analogue of Fig.~\ref{fig.fires-supp} for powerlaw model.  
    (a) Log-binned empirical distribution of fire magnitude (circles) and the maximum likelihood fit to a powerlaw model (dotted black line) of values above the lower cutoff. Inset: Expectations of sample maximum estimated as the mean over $10^5$ independent samples of length $N$ downsampled (i.e., drawn i.i.d. without replacement) either from the original data or surrogates produced from the full dataset. (b) Rates of rejection estimated from 1000 hypothesis tests, each using a sample of length $N$ from the observed dataset, 99 surrogates, and either the KS distance or maximum. Only values of the original data which equal or exceed the fitted lower cutoff $x_{\min} = 40.5\ [500 \textrm{m}^2]$ are considered.
    }
    \label{fig.fires-powerlaw-supp}
\end{figure}

\begin{figure}
\includegraphics[width=0.9\columnwidth]{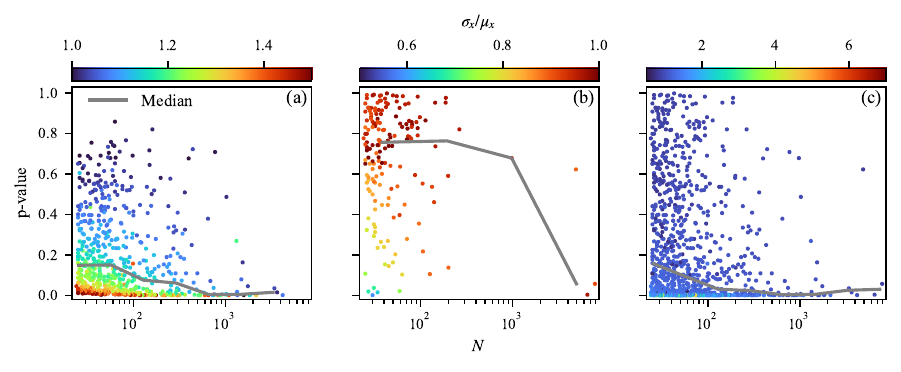}
    \caption{Analogues of inset of Fig.~\ref{fig.words} when considering other ranges of burstiness $\sigma_x/\mu_x$. Constrained surrogates to test the distribution of words in texts. Reported is the p-value returned from tests of an exponential distribution of waiting times $x$ between repetitions of the same word in the book \emph{The Voyage of the Beagle}, by Charles Darwin. (a) 614 considered words for which $1 \leq \sigma_x/\mu_x < 1.5$. (b) 169 considered words for which $0.5 \leq \sigma_x/\mu_x < 1$. (c) All 1000 considered words. Each marker corresponds to a word, and its color indicates its corresponding coefficient of variation $\sigma_x/\mu_x$. The median p-value is shown as a function of $N$ (gray line).
    Hypothesis tests use 999 constrained surrogates and employ $\sigma_x/\mu_x$ as discriminating statistic.
    We considered the the 1000 most frequent words and used lower cutoff $x_{\min} = 9.5$, which approximates the typical length of a clause in written English~\cite{lyu2022marked}.}
    \label{fig.words-p-vs-N}
\end{figure}

\begin{table*}[htbp]
\caption{
Constrained and typical powerlaw surrogates lead to the same conclusions about the validity of different models for city population datasets. The p-values identified via tests using KS-distance and $999$ surrogates representing the hypothesis that the $N$ elements of a dataset which exceed the lower cut-off $\hat{x}_{\min}$ arose under an i.i.d. (from left to right): powerlaw, lognormal, exponential, truncated Gaussian, and uniform process. Values which, ignoring issues associated with multiple tests, would allow rejection at the 10\% level of significance~\cite{clauset2009power} are shown in \textit{italics}. The lower cut-off $x_{\min}$ is chosen as 0.5 less than the minimum observed population of a city with at least 10\ 000 inhabitants (see Ref.~\cite{altmann2025statistical}). 
The number of elements $N$ in the fitted dataset is also shown. Data from \protect\url{https://github.com/edugalt/StatisticalLaws/tree/main/data/cities/}.}
\small
\label{tab.cities}
\centering
\begin{ruledtabular}
\begin{tabular}{ccccccccc}
Country & \multicolumn{2}{c}{Dataset} & Surrogate & & & Model & & \\
\cline{2-3}\cline{5-9}
 & $x_{\min}$ & $N$ & & Powerlaw & Lognormal & Exponential & Truncated Gaussian & Uniform \\
 \hline
Australia & 10\ 544.5 & 102 & Constrained & 0.35 & 0.34 & \textit{0.0005} & \textit{0.0005} & \textit{0.0005} \\
 & & & Typical & 0.37 & 0.36 & \textit{0.0005} & \textit{0.0005} & \textit{0.0005} \\
Brazil & 9999.5 & 3058 & Constrained & \textit{0.0005} & \textit{0.0005} & \textit{0.0005} & \textit{0.0005} & \textit{0.0005} \\
 & & & Typical & \textit{0.0005} & \textit{0.0005} & \textit{0.0005} & \textit{0.0005} & \textit{0.0005} \\
Germany & 33\ 741.5 & 107 & Constrained & \textit{0.0005} & 0.59 & \textit{0.0095} & 0.12 & \textit{0.0005} \\
 & & & Typical & \textit{0.0005} & 0.60 & \textit{0.0005} & 0.24 & \textit{0.0005} \\
UK & 50\ 029.5 & 100 & Constrained & \textit{0.058} & 0.32 & \textit{0.0005} & \textit{0.0015} & \textit{0.0005} \\
 & & & Typical & \textit{0.060} & 0.34 & \textit{0.0005} & \textit{0.0005} & \textit{0.0005} \\
USA & 54\ 060.5 & 381 & Constrained & \textit{0.0005} & \textit{0.0005} & \textit{0.0005} & \textit{0.0005} & \textit{0.0005} \\
 & & & Typical & \textit{0.0005} & \textit{0.0005} & \textit{0.0005} & \textit{0.0005} & \textit{0.0005}
\end{tabular}
\end{ruledtabular}
\end{table*}

\begin{figure}
\includegraphics[width=0.99\linewidth]{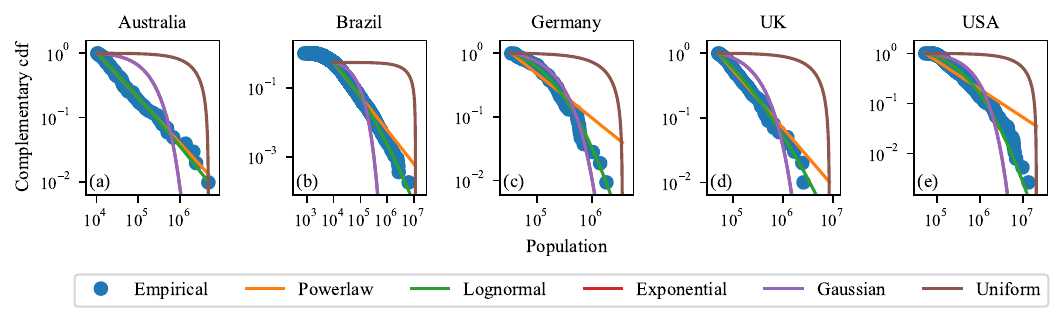}
    \caption{%
        Fits of empirical city population data (blue) within a country to five different models: powerlaw (orange), lognormal (green), exponential (red), truncated Gaussian (purple), and uniform (brown). Five different countries are considered: (a) Australia, (b) Brazil, (c) Germany, (d) UK, and (e) USA. For a given country, lower cut-off $x_{\min}$ is the same for each model (see Table~\ref{tab.cities}).
    }
    \label{fig.cities-fits}
\end{figure}

\begin{figure*}[h]
    \includegraphics[width=\textwidth]{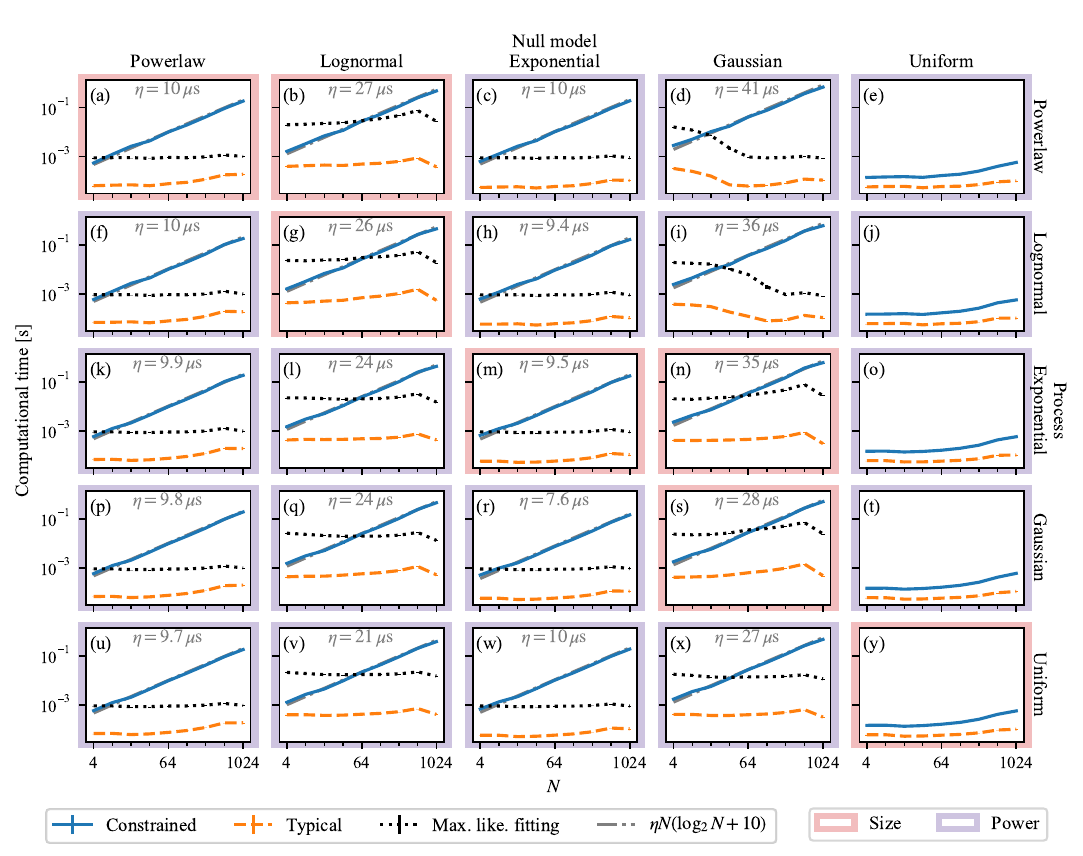}
    \caption{The computational cost  of constrained surrogates is roughly proportional to the number of transitions used. Computational cost per surrogate estimated as the average over $1,000$ hypothesis tests, each using an independent sample of length $N$ i.i.d. from a (from top to bottom): powerlaw; lognormal; exponential; truncated Gaussian; and uniform distribution. Each test uses 19 constrained (blue solid) or typical (orange solid) surrogates representing the null hypothesis of a (from left to right): (a-e) powerlaw; (f-j) lognormal; (k-o) exponential; (p-t) truncated Gaussian; and (u-y) uniform distribution. The time required to estimate maximum likelihood parameters is shown separately (black dotted) because this cost is incurred once per input sequence to generate typical surrogates (but not constrained surrogates) and once per typical surrogate (each of which has a distinct maximum likelihood parameter) when calculating certain statistics (e.g., the KS distance). With $2\tilde{r} = N \left(\log_2 N + 10\right)$ transitions per constrained surrogate, we fit the computational time $T=\eta \times 2\tilde{r}$ (gray dot-dashed) by least squares in $\log T$, with the fitted time per transition $\eta$ reported in the corresponding panel. Panels showing tests for which the null hypothesis is correct (i.e., for which the rate of rejection is the size of the test) have red borders, while panels showing tests for which the null hypothesis is incorrect (i.e., for which the rate of rejection is the test's power to reject an incorrect hypothesis) are given purple borders. In this plot, each constrained surrogate is generated independently from the original observed sequence instead of from the previously generated constrained surrogate.
    }
    \label{fig.comp-cost-vs-N}
\end{figure*}

\end{document}